# The Damping and Excitation of Galactic Warps by Dynamical Friction

Robert W. Nelson and Scott Tremaine
Canadian Institute for Theoretical Astrophysics

## ABSTRACT

We investigate the dynamical interaction of galactic warps with the surrounding dark matter halo, using analytic perturbation theory. A precessing warp induces a density wake in the collisionless dark matter, which acts back on the original warp, transferring energy and angular momentum between the warp and halo (dynamical friction). In most cases dynamical friction damps the warp, but in unusual circumstances (such as a halo that rotates in the same direction as the precession of the warp, or a warp in the equatorial plane of an axisymmetric prolate halo) friction can excite a warp. The damping/excitation time is usually short compared to the Hubble time for realistic systems. Thus most warps cannot be primordial; they must be maintained by some ongoing excitation mechanism.

## 1  Introduction

The origin and maintenance of the 'integral sign' warps commonly seen in the outer parts of disk galaxies is a longstanding puzzle in galactic dynamics (for a recent review see Binney 1992). Many if not most disk galaxies—even isolated ones—appear to be warped, which implies that warps are either long-lived or repeatedly excited (Sánchez-Saavedra et al. 1990, Bosma 1991). Following an initial suggestion by Lynden-Bell (1965), Hunter & Toomre (1969; hereafter HT) explored the possibility that warps are discrete modes of oscillation of a thin, self-gravitating disk. They found, however, that isolated disks with realistic density profiles actually support a continuum of bending modes; consequently, any packet of bending waves should disperse. Dekel & Shlosman (1983) and Toomre (1983) suggested a solution to this problem, which was analyzed in detail by Sparke & Casertano (1988; hereafter SC): realistic disks can support discrete bending modes if they are embedded in the static potential of a flattened dark matter halo—these modes are distorted versions of the neutral tilt mode of an isolated disk. Thus the gravitational influence of galaxy halos may play an important role in the maintenance of galactic warps; conversely, by studying the dynamics of warps in galaxies we may learn about the shape and dynamics of the unseen halos.

Unfortunately, galaxy halos are not simply static potentials: the collisionless halo material—whatever its nature—must respond to the gravitational field of an embedded disk. A warped, precessing disk induces a density wake in the dark matter halo and the gravitational field from the wake will then act back on the disk; in this manner energy and angular momentum are transferred between disk and halo in a manner analogous to the Landau damping of waves in a collisionless plasma or dynamical friction on a point mass travelling through a stellar system. In general, then, any bending mode of a galaxy disk is not precisely neutral; the question is whether the dynamical coupling between the halo and the disk is strong enough to damp—or excite, as we shall see—bending modes of a galaxy on timescales less than the age of the galaxy itself[1].

This issue was first investigated by Bertin & Mark (1980), who approximated the warp as a tightly-wound (WKB) bending wave, embedded in an infinite, homogeneous, and isotropic 'halo' having a Maxwell-Boltzmann velocity distribution. Their analysis predicted that the amplitude of the bending wave would grow inside the corotation radius—and on a short timescale. Unfortunately, this interesting conclusion is suspect, because Bertin and Mark neglected the unperturbed halo's contribution to the vertical restoring force (see Section 3.1).

In this paper we examine the role of a responsive halo in damping or exciting bending disturbances in galactic disks. We extend Bertin and Mark's seminal analysis to include warps and halos with more realistic shapes, as well as halos with anisotropic velocity distribution and net rotation. In particular, if the orbits of halo particles—we shall refer to these as 'stars' although their nature is unknown—are quasiperiodic, energy and angular momentum exchange between disk and halo occurs via stars whose orbits are nearly resonant with the frequency of the time-varying disk potential (Tremaine & Weinberg 1984). In this situation, the direction of energy and angular momentum flow can be from disk to halo or vice versa, so that dynamical friction with the halo may either damp or excite warps, depending on the details of the halo structure.

Our analysis of this phenomenon is more elaborate than any in the literature so far, but still has a number of limitations. We do not consider the self-gravity of the halo response. We calculate the unperturbed orbits of stars in the disk plus halo system by assuming that the potential is spherical, and assume that the halo is axisymmetric for the purpose of computing the precession rate and energy of the bending disturbance; whereas cosmological N-body simulations of the collapse of protogalactic density peaks typically result in triaxial halos. We use linear perturbation theory, whose validity is suspect since warps often have substantial amplitudes.

Despite these shortcomings, we believe that the principal results of our analysis are robust. In particular, we find that dynamical friction from the halo can either damp or excite bending disturbances in a disk, although damping is the most common outcome for plausible disk/halo configurations. The characteristic growth or decay time is generally less than a Hubble time, implying that dynamical friction plays a major role in determining the present properties of galaxy warps.

In Section 2 we review the dynamics of bending disturbances in a thin disk embedded in a *static* halo potential. In Section 3 we relax this assumption and consider the coupling of the bending perturbation to the dynamics of the halo stars. In particular, we give general formulae for the damping/excitation rate of bending disturbances from dynamical friction with the halo. In Section 4 we present numerical calculations of damping/excitation times for reasonably realistic halo and disk potentials and warp shapes. Finally, in Section 5 we discuss the implications of our results.

---

[1] 'One key question about this kind of proposal is how long the central disk can thereafter remain tilted despite dynamical friction against the doubtless flabby halo' (Toomre 1983).

## 2  Warp Dynamics

We review the behavior of linear bending waves of a zero-thickness, axisymmetric, self-gravitating disk embedded in a static halo potential, delaying until Section 3 consideration of the coupling of bending waves to a responsive halo. Many, but not all, of the results we describe are presented in HT, Sparke (1984), and SC.

### 2.1  The equation of motion

We work in cylindrical coordinates $(R, \phi, z)$ and take the unperturbed disk to lie in the $z = 0$ plane, so that position in the disk is described by the vector $\mathbf{R} \equiv (R, \phi)$. We consider a razor-thin axisymmetric disk of surface density $\mu(R)$, rotating at angular speed $\Omega(R) > 0$. The equation of motion in the inertial frame for small vertical displacements, $Z(R, \phi, t)$, is (HT)

$$\frac{D^2 Z}{Dt^2} \equiv \left( \frac{\partial}{\partial t} + \Omega \frac{\partial}{\partial \phi} \right)^2 Z = F_h + F_{sg} + F_{ext}, \tag{1}$$

where the three terms on the right side are the vertical components of the force per unit mass arising from the gravitational field of the static halo ($F_h$) and the disk ($F_{sg}$), and from any other sources ($F_{ext}$). We assume that the halo is axisymmetric and that its density is smooth near the disk, so that the halo potential can be expanded in a Taylor series

$$U_h(R, z) = U_h(R, 0) + \tfrac{1}{2} \nu_h^2(R) z^2, \quad \text{where} \quad \nu_h^2(R) = \frac{\partial^2 U_h}{\partial z^2}(R, z = 0) \tag{2}$$

is the square of the frequency of small vertical oscillations in the field of the halo. Thus

$$F_h(\mathbf{R}, t) = -\nu_h^2(R) Z(\mathbf{R}, t). \tag{3}$$

It is sometimes useful to decompose the radial force $-\Omega^2 R$ into its contributions from the halo and the disk,

$$\Omega^2(R) = \Omega_h^2(R) + \Omega_d^2(R), \tag{4}$$

where

$$\Omega_h^2(R) = \frac{1}{R} \frac{\partial U_h}{\partial R}(R, z = 0), \qquad \Omega_d^2(R) = -\frac{G}{R} \frac{d}{dR} \int \frac{d\mathbf{R}' \mu(R')}{|\mathbf{R} - \mathbf{R}'|}. \tag{5}$$

The vertical restoring force from a slightly distorted disk is given by HT,

$$F_{sg}(\mathbf{R}, t) = -G \int d\mathbf{R}' \mu(R') \frac{[Z(\mathbf{R}, t) - Z(\mathbf{R}', t)]}{|\mathbf{R} - \mathbf{R}'|^3}. \tag{6}$$

We may write the disturbance $Z$ as a superposition of eigenfunctions of the form

$$Z(\mathbf{R}, t) = \mathrm{Re}\left\{ h(R)e^{i(m\phi - \omega t)} \right\}, \tag{7}$$

where we assume $m \geq 0$ without loss of generality. If $\omega$ is real and $m \neq 0$, the pattern speed $\Omega_p \equiv \omega/m$ is the angular speed of the rotating frame in which the disturbance is stationary.

When external forces are absent ($F_{ext} = 0$), the dynamical equation (1) for a single eigenfunction reduces to

$$\left\{ [\omega - m\Omega(R)]^2 - \nu_h^2(R) \right\} h(R) = G \int_0^\infty dR' R' \mu(R')[h(R)H(R, R') - h(R')I_m(R, R')] \tag{8}$$

where[2]

$$H(R, R') = \int_0^{2\pi} \frac{d\psi}{(R^2 + R'^2 - 2RR'\cos\psi)^{3/2}}, \quad I_m(R, R') = \int_0^{2\pi} \frac{d\psi \cos m\psi}{(R^2 + R'^2 - 2RR'\cos\psi)^{3/2}}; \tag{9}$$

these are Laplace coefficients and can be evaluated in terms of elliptic integrals.

The component of the torque on the disk parallel to the $z = 0$ plane may be written $\boldsymbol{\tau} = \tau_x \mathbf{e}_x + \tau_y \mathbf{e}_y$, where

$$\tau_x + i\tau_y = \int d\mathbf{R} \mu(R)[F_z(y - ix) - z(F_y - iF_x)]; \tag{10}$$

here $\mathbf{F} = (F_x, F_y, F_z)$ is the force per unit mass on the disk at $\mathbf{R}$, composed of a radial component $-\Omega^2 R$ and a vertical component $-\nu_h^2 Z + F_{sg}$. Thus

$$\tau_x + i\tau_y = -i \int d\mathbf{R} \mu(R) Re^{i\phi}[(\Omega^2 - \nu_h^2)Z(\mathbf{R}, t) + F_{sg}]. \tag{11}$$

Since the torque from the disk on itself must be zero we have (Kuijken 1991)

$$0 = \int d\mathbf{R} \mu(R) Re^{i\phi}[\Omega_d^2 Z(\mathbf{R}, t) + F_{sg}]; \tag{12}$$

(a result which can also be derived by manipulating eqs. 5 and 6) and

$$\tau_x + i\tau_y = -i \int d\mathbf{R} \mu(R) Re^{i\phi}(\Omega_h^2 - \nu_h^2)Z(\mathbf{R}, t). \tag{13}$$

---

[2]The functions $H$ and $I_m$ diverge as $R' \to R$ but the integral in equation (8) is nevertheless well-defined.

These equations degenerate to $0 = 0$ when $m \neq 1$ but when $m = 1$ we may use equations (1) and (7) to reduce equation (12) to the form

$$\int_0^\infty dR\, R^2 \mu(R)[\Omega_p^2 - 2\Omega\Omega_p + (\Omega_h^2 - \nu_h^2)]h(R) = 0. \tag{14}$$

This constraint provides a quadratic equation for the pattern speed $\Omega_p$ in terms of the eigenfunction $h(R)$; its unique feature is that the self-gravity of the disk does not have to be evaluated explicitly. In general, one of the roots of equation (14) is spurious and we shall assume that the correct root is the one with the smallest absolute value. The motivation for this choice is that we focus on the distorted tilt mode, which has a small pattern speed.

Numerical solutions of the eigenvalue equation (8) for various disk and halo density profiles are given by HT, Sparke (1984) and SC. The following special cases offer useful insight:

1. In the absence of disk self-gravity, equation (8) reduces to

$$\omega = m\Omega(R) \pm \nu_h(R), \tag{15}$$

   as does (14) when $m = 1$. In general, the combination of terms on the right side will depend on radius, so the eigenfunctions are singular (van Kampen modes), and any spatially extended initial disturbance will damp by phase mixing.

2. If the halo is spherical, the disk is neutrally stable to tilting. To see this from equation (8) we use equations (5) and (9) to write

$$\Omega_d^2(R) = \frac{G}{R} \int_0^\infty dR'\, R'\mu(R')[RH(R,R') - R'I_1(R,R')]; \tag{16}$$

   it follows that when $m = 1$ and the halo is spherical ($\nu_h = \Omega_h$) a solution of (8) is $\omega = 0$, $h(R) \propto R$ (the 'tilt mode').

3. For a nearly spherical halo, the tilt mode is distorted and acquires a small but non-zero pattern speed. In equation (14) we may therefore drop the term proportional to $\Omega_p^2$; moreover, in a first approximation we can replace the eigenfunction $h(R)$ by the tilt mode $h(R) \propto R$. Thus we obtain (SC)

$$\Omega_p \simeq \frac{\int_0^\infty (\Omega_h^2 - \nu_h^2)\mu(R)R^3 dR}{2\int_0^\infty \Omega(R)\mu(R)R^3 dR}; \tag{17}$$

   note that the angular frequency in the denominator includes the contribution from both disk and halo, while the frequencies in the numerator represent only the halo contribution. Equation (17) simply says that the precession rate of the warped disk is the torque from the halo divided by the off-axis angular momentum (Kuijken 1991); we shall refer to it as the Sparke-Casertano formula for the pattern speed.

The distorted tilt mode is closely related to the Laplace invariable plane[3]: each annulus of the disk is in the Laplace plane determined by the halo and the other annuli. Thus we shall refer to this mode as the 'Laplace mode' (SC use the term 'modified tilt mode').

## 2.2 Energy and angular momentum

To determine the energy $E_w$ associated with a bending disturbance $Z(\mathbf{R}, t)$, we assume that the disturbance has been generated by an external potential $U_{ext}(\mathbf{r}, t)$, where $\mathbf{r} = (R, \phi, z) = (\mathbf{R}, z)$. If the disk has density $\rho(\mathbf{r}, t)$ and velocity $\mathbf{v}(\mathbf{r}, t)$ then the rate at which the external potential does work on the disk is

$$\frac{dE_w}{dt} = \int d\mathbf{r}\, \rho \mathbf{v} \cdot \mathbf{F}_{ext} = -\int d\mathbf{r}\, \rho \mathbf{v} \cdot \boldsymbol{\nabla} U_{ext}. \tag{18}$$

We may write $\mathbf{v} = (0, \Omega R, DZ/Dt)$; the radial and azimuthal velocities have been set to their unperturbed values since we are working to first order in the small quantity $Z(\mathbf{R}, t)$ and to this order the horizontal and vertical motions are decoupled. Thus

$$
\begin{aligned}
\frac{dE_w}{dt} &= -\int d\mathbf{r}\, \rho \left[\Omega U_{ext,\phi} + (Z_{,t} + \Omega Z_{,\phi}) U_{ext,z}\right], \\
&= \int d\mathbf{r}\, \left[\Omega \rho_{,\phi} U_{ext} - \rho Z_{,t} U_{ext,z} + \Omega \rho_{,z} Z_{,\phi} U_{ext}\right],
\end{aligned} \tag{19}
$$

where the second line follows from integration by parts. The density of our razor-thin disks may be written $\rho(\mathbf{r}, t) = \mu(R)\delta[z - Z(\mathbf{R}, t)]$. Thus

$$\rho_{,\phi} = -\mu \delta'(z - Z) Z_{,\phi} = -\rho_{,z} Z_{,\phi}\,. \tag{20}$$

Substituting this result in equation (19) we find that the first and third terms cancel, so that

$$\frac{dE_w}{dt} = -\int d\mathbf{r}\, \rho Z_{,t} U_{ext,z} = \int d\mathbf{R}\, \mu F_{ext} Z_{,t} \tag{21}$$

where as usual $F_{ext}$ is the $z$-component of the external force (note that the expression is not $\int d\mathbf{R}\, \mu F_{ext}(DZ/Dt)$, as one might expect from considering only vertical motions).

We may eliminate the external force $F_{ext}$ from equation (21) using equations (1), (3) and (6). Then using integration by parts with respect to $\phi$, we can convert the right side of equation (21) into the total time derivative of a quantity which must equal the energy $E_w$ to within a constant. Setting this constant to zero when $Z = 0$, we obtain

$$E_w = \frac{1}{2} \int d\mathbf{R} \mu(R)[Z_{,t}^2 + \nu_h^2 Z^2 - \Omega^2 Z_{,\phi}^2] + \frac{G}{4} \int d\mathbf{R} d\mathbf{R}' \mu(R)\mu(R') \frac{[Z(\mathbf{R},t) - Z(\mathbf{R}',t)]^2}{|\mathbf{R} - \mathbf{R}'|^3}. \quad (22)$$

---

[3]Consider a flattened planet that is accompanied by a massive satellite on an inclined orbit. A test particle orbiting the planet in the Laplace plane will remain in that plane (except for short-period oscillations) despite the perturbations from the planet's equatorial bulge and the massive satellite. The normal to the Laplace plane lies in the plane determined by the spin and orbital angular momentum vectors of the planet and satellite, while the orientation of the plane is fixed so that the angular momentum vectors of the planet, satellite and test particle all precess at the same rate.

In the absence of an external force, the energy (22) is conserved[4].

Note that the energy of a disturbance is not guaranteed to be positive, since the first integrand can be negative. HT examined the sign of the energy for azimuthal wavenumbers $m = 0$ (axisymmetric) and $m = 1$ (tilt or warp). When $m = 0$ $Z_{,\phi} = 0$ so equation (22) implies that the energy is positive (or zero, in the trivial case $\nu_h = 0$, $Z =$ constant, corresponding to the translation of an isolated disk). Since unstable modes must have zero energy, all $m = 0$ modes are stable.

In the case $m = 1$ it is useful to decompose the angular speed into contributions from the disk and halo (eqs. 4, 5). HT then show that

$$-\frac{1}{2} \int d\mathbf{R} \mu(R)\Omega_d^2(R)Z_{,\phi}^2 + \frac{G}{4} \int d\mathbf{R} d\mathbf{R}' \mu(R)\mu(R') \frac{[Z(\mathbf{R},t) - Z(\mathbf{R}',t)]^2}{|\mathbf{R} - \mathbf{R}'|^3}$$

$$= \frac{G}{4} \int dR dR' d\phi \mu(R)\mu(R')[RZ(R',\phi,t) - R'Z(R,\phi,t)]^2 I_1(R,R'), \quad (24)$$

where $I_m$ is defined in equation (9). Thus the energy of an $m = 1$ disturbance may be written

$$E_w = \frac{1}{2} \int d\mathbf{R} \mu(R)[Z_{,t}^2 + (\nu_h^2 - \Omega_h^2)Z^2] + \frac{G}{4} \int dR dR' d\phi \mu(R)\mu(R')[RZ(R',\phi,t) - R'Z(R,\phi,t)]^2 I_1(R,R'). \quad (25)$$

If there is no halo ($\nu_h = \Omega_h = 0$) the energy is positive or zero; zero energy corresponds to the disturbance $Z = AR\cos(\phi - \phi_0)$ where $A$ and $\phi_0$ are constants (physically, an isolated disk is neutrally stable to a uniform tilt). Since unstable modes must have zero energy, all $m = 1$ modes of an isolated disk are stable, a conclusion due to HT.

When the disk is embedded in a halo, $m = 1$ disturbances can have positive or negative energy. We shall call an axisymmetric halo 'vertically stiff' if $\nu_h > \Omega_h$ at all radii (which is generally the case if the disk lies in the equatorial plane of an oblate mass distribution), and 'vertically soft' if $\nu_h < \Omega_h$ (which occurs for prolate distributions). Equation (25) shows that the total energy of an $m = 1$ disturbance is positive—so that $m = 1$ modes are stable—whenever the halo is vertically stiff.

---

[4]This result for the energy is perhaps somewhat unexpected: in the absence of self-gravity, the most natural way to derive an energy is to treat the disk particles as independent vertical harmonic oscillators. This procedure yields

$$E_w' = \frac{1}{2} \int d\mathbf{R} \mu(R)[(DZ/Dt)^2 + \nu_h^2 Z^2] = \frac{1}{2} \int d\mathbf{R} \mu(R)(Z_{,t}^2 + \nu_h^2 Z^2 + \Omega^2 Z_{,\phi}^2 + 2\Omega Z_{,t} Z_{,\phi}), \quad (23)$$

which is not the same as the first term of (22). Both $E_w$ and $E_w'$ are conserved if $F_{sg} = F_{ext} = 0$, but equation (22) is the only energy-like quantity conserved when self-gravity is present (except in the special case of uniform rotation, which is described at the end of this section).

The expression for the energy (22) can be rewritten in a simpler form. We operate on equation (1) with $\int d\mathbf{R} \mu(R)$; setting $F_{ext} = 0$ we obtain

$$\int d\mathbf{R} \mu(R)(ZZ_{,tt} + 2\Omega ZZ_{,t\phi} \;\; + \;\; \Omega^2 ZZ_{,\phi\phi} + \nu_h^2 Z^2)$$
$$= -G \int d\mathbf{R} d\mathbf{R}' \mu(R)\mu(R') \frac{Z(\mathbf{R},t)[Z(\mathbf{R},t) - Z(\mathbf{R}',t)]}{|\mathbf{R} - \mathbf{R}'|^3}$$
$$= -\frac{G}{2} \int d\mathbf{R} d\mathbf{R}' \mu(R)\mu(R') \frac{[Z(\mathbf{R},t) - Z(\mathbf{R}',t)]^2}{|\mathbf{R} - \mathbf{R}'|^3}, \tag{26}$$

where the last expression is obtained from the second by interchanging primed and unprimed variables and averaging with the original expression. Using this result to eliminate the last term from equation (22) we find

$$E_w = \tfrac{1}{2} \int d\mathbf{R} \mu(R)[Z_{,t}^2 - ZZ_{,tt} - 2\Omega ZZ_{,t\phi} - \Omega^2 Z_{,\phi}^2 - \Omega^2 ZZ_{,\phi\phi}]. \tag{27}$$

The terms proportional to $\Omega^2$ cancel upon integrating by parts with respect to $\phi$. Thus

$$E_w = \tfrac{1}{2} \int d\mathbf{R} \mu(R)[Z_{,t}^2 - ZZ_{,tt} - 2\Omega ZZ_{,t\phi}]. \tag{28}$$

For a disturbance of the form (7) with $m \neq 0$ and real frequency $\omega = m\Omega_p$, we have

$$E_w = -\pi m^2 \Omega_p \int dR R \mu(R)|h(R)|^2 (\Omega - \Omega_p). \tag{29}$$

Thus disturbances with zero pattern speed have zero energy; if the pattern speed is negative (retrograde precession) the energy is always positive; if the pattern speed is small and positive the energy is negative.

We can also work out the $z$-component of angular momentum associated with a disturbance excited by an external potential:

$$\frac{dJ_w}{dt} = -\int d\mathbf{r} \rho U_{ext,\phi}$$
$$= \int d\mathbf{r} \rho_{,\phi} U_{ext}$$
$$= -\int d\mathbf{r} \rho_{,z} Z_{,\phi} U_{ext}$$
$$= \int d\mathbf{r} \rho U_{ext,z} Z_{,\phi}, \tag{30}$$

where the third line follows from equation (20). We eliminate $U_{ext,z} = -F_{ext}$ using equations (1) and (3) and derive an expression for the angular momentum by a similar procedure to the one used to derive equation (22). In this derivation we use the results

$$Z_{,\phi} Z_{,tt} = (Z_{,\phi} Z_{,t})_{,t} - \tfrac{1}{2}(Z_{,t}^2)_{,\phi} \tag{31}$$

and

$$
\begin{aligned}
\int d\mathbf{R}\mu(R)Z_{,\phi}\,F_{sg} &= -G\int d\mathbf{R}d\mathbf{R}'\mu(R)\mu(R')\frac{Z_{,\phi}\,(Z-Z')}{|\mathbf{R}-\mathbf{R}'|^3}\\
&= -\frac{G}{2}\int d\mathbf{R}d\mathbf{R}'\mu(R)\mu(R')\frac{[(Z-Z')^2]_{,\phi}}{|\mathbf{R}-\mathbf{R}'|^3}\\
&= \frac{G}{2}\int d\mathbf{R}d\mathbf{R}'\mu(R)\mu(R')(Z-Z')^2\frac{\partial}{\partial\phi}|\mathbf{R}-\mathbf{R}'|^{-3},
\end{aligned}
\tag{32}
$$

which is zero, because the integrand to the left of $\partial/\partial\phi$ is symmetric under interchange of primed and unprimed variables, while $\partial|\mathbf{R}-\mathbf{R}'|^{-3}/\partial\phi$ is antisymmetric.

In this way we find

$$
J_w = -\int d\mathbf{R}\mu(R)(Z_{,t}\,Z_{,\phi}+\Omega Z_{,\phi}^2).
\tag{33}
$$

Thus, in the absence of external forces, there are two independent conserved quantities that are quadratic in the disturbance strength. These quantities correspond to the energy $E_w$ and $z$-component of angular momentum $J_w$. For a disturbance of the form (7) with real frequency $\omega=m\Omega_p$, we have

$$
J_w = -\pi m^2\int dR\,R\mu(R)|h(R)|^2(\Omega-\Omega_p).
\tag{34}
$$

Comparison of equation (34) with equation (29) shows that

$$
E_w = \Omega_p J_w.
\tag{35}
$$

We close this section by considering the special case of disks with constant angular speed, $\Omega(R)=\Omega_0=$const. Combining equations (22) and (33) we obtain the conserved quantity

$$
E_w-\Omega_0 J_w = \tfrac{1}{2}\int d\mathbf{R}\mu(R)[(Z_{,t}+\Omega_0 Z_{,\phi})^2+\nu_h^2 Z^2]+\frac{G}{4}\int d\mathbf{R}d\mathbf{R}'\mu(R)\mu(R')\frac{[Z(\mathbf{R},t)-Z(\mathbf{R}',t)]^2}{|\mathbf{R}-\mathbf{R}'|^3}.
\tag{36}
$$

Since this quantity is non-negative, all bending disturbances in a uniformly rotating disk are stable.

## 2.3 The WKB approximation

The equation of motion (1) for bending waves can be solved analytically in the WKB or short-wavelength limit. This solution offers useful insights, even though the WKB approximation is poor for the bending waves actually seen in galaxies.

In the WKB limit the radial wavelength $\lambda$ is much shorter than any relevant radial scale length, i.e.,

$$
\lambda\ll R,\qquad \lambda\ll\left|\frac{d\ln\mu(R)}{dR}\right|^{-1},
\tag{37}
$$

but long compared to the wave amplitude,

$$\lambda \gg |Z|. \tag{38}$$

We write the vertical disturbance as

$$Z(\mathbf{R}, t) = \text{Re} \left\{ A(R, t) e^{i \int^R K(R) dR} e^{i(m\phi - \omega t)} \right\}, \tag{39}$$

where $K(R) = 2\pi/\lambda(R)$ is the local wavenumber and $A(R, t)$ is a slowly varying amplitude.

The vertical gravitational force from the disturbed disk (eq. 6) can be evaluated using local Cartesian coordinates, $\mathbf{R}' = \mathbf{R} + x\mathbf{e}_x + y\mathbf{e}_y$, where at $\mathbf{R}$ the $x$–axis points radially outward and the $y$–axis points azimuthally. Because the radial wavelength is short, we neglect the radial variation in the surface density $\mu(R')$ and the azimuthal variation in $Z$, and write

$$Z(\mathbf{R}', t) = \text{Re}[A_0 e^{i(Kx - \omega t)}], \qquad Z(\mathbf{R}, t) = \text{Re}[A_0 e^{-i\omega t}], \tag{40}$$

where

$$A_0 \equiv A(R, t) e^{i \int^R K(R) dR} e^{im\phi(\mathbf{R})}. \tag{41}$$

Then equation (6) yields

$$
\begin{aligned}
F_{sg}(\mathbf{R}, t) &= -G\mu(R) \, \text{Re} \left\{ A_0 e^{-i\omega t} \int dx \, dy \frac{[1 - \exp(iKx)]}{(x^2 + y^2)^{3/2}} \right\} \\
&= -4G\mu(R) |K| Z(\mathbf{R}, t) \int_0^\infty \frac{dv}{(1 + v^2)^{3/2}} \int_0^\infty du \frac{(1 - \cos u)}{u^2} \\
&= -2\pi G\mu(R) |K| Z(\mathbf{R}, t).
\end{aligned}
\tag{42}
$$

The WKB dispersion relation follows from the dynamical equation (1),

$$D(\omega, K) \equiv [\omega - m\Omega(R)]^2 - 2\pi G\mu(R)|K| - \nu_h^2(R) = 0. \tag{43}$$

At the corotation radius, the angular speed of the disk equals the pattern speed, $\Omega(R) = \Omega_p = \omega/m$. At the vertical resonance radii $K(R) = 0$, that is, $\omega = m\Omega(R) \pm \nu_h(R)$ (*cf.* eq. 15). Because the dispersion relation can only be satisfied when $m^2(\Omega - \Omega_p)^2 \geq \nu_h^2$, there is a forbidden region between the two vertical resonance radii and including corotation, in which WKB bending waves cannot propagate.

The radial group velocity for WKB bending waves is (Toomre 1969, Whitham 1974)

$$c_g = \frac{\partial \omega}{\partial K} = -\frac{\pi G\mu(R) \, \text{sgn}(K)}{m[\Omega(R) - \Omega_p]}. \tag{44}$$

Thus the direction of propagation depends on whether the waves are leading ($K < 0$) or trailing ($K > 0$), and on whether they are inside corotation ($\Omega > \Omega_p$, assuming that the angular speed decreases outward) or outside.

The equation of motion for vertical disturbances (1) can be derived from a Lagrangian. The dynamics in the WKB limit are therefore described by an averaged Lagrangian that is obtained by averaging over the short-wavelength oscillations (Whitham 1974),

$$L(\omega, K, A) = \tfrac{1}{2}\pi R \mu |A|^2 D(\omega, K),$$  (45)

where $|A|$ is the wave amplitude and $D(\omega, K)$ is the dispersion relation (43). The energy density of the wave per unit radius is

$$\epsilon(R) = \omega \frac{\partial L}{\partial \omega} = -\pi m^2 R \mu(R) \Omega_p (\Omega - \Omega_p) |A|^2.$$  (46)

The energy density is positive if the pattern speed is prograde ($\Omega_p > 0$) and outside corotation ($\Omega < \Omega_p$), or if the pattern speed is retrograde. The total energy of the disturbance is

$$E_w = \int \epsilon(R) dR,$$  (47)

which is the same as the energy derived in the previous section (eq. 29)—in this case it happens that the WKB approximation is exact.

The density of $z-$angular momentum per unit radius is (Bertin & Mark 1980)

$$j(R) = m \frac{\partial L}{\partial \omega} = \frac{\epsilon(R)}{\Omega_p},$$  (48)

which is exactly consistent with equation (34).

## 3  Dynamical Friction Between a Warped Disk and a Spherical Halo

So far we have treated the halo potential as a static background field, which influences the disk only through its contribution to the vertical and radial restoring force. In fact, the halo is composed of stars whose orbits respond to the time-varying gravitational field of the warped disk. The gravitational attraction between the coherent response of the halo and the warped disk transfers energy and angular momentum between the two components, thereby damping—or possibly exciting—the warp (dynamical friction).

To describe this response, we write the halo potential as $U_h(\mathbf{r}) = U_h^0(\mathbf{r}) + U_h^1(\mathbf{r}, t)$, where the two terms denote the static, unperturbed halo potential and the potential arising from its response to the warp. The corresponding densities are $\rho_h^0(\mathbf{r})$ and $\rho_h^1(\mathbf{r}, t)$. For consistency with Section 2.2, we define the energy of the bending disturbance in the disk, $E_w$, to include the kinetic and self-gravitational energies of the disk material plus its potential energy in the static field $U_h^0$, but *not* its potential energy in the response field $U_h^1$; the forces arising from the latter field are treated as an external force $\mathbf{F}_{ext} = -\boldsymbol{\nabla} U_h^1$. Thus

$$
\begin{aligned}
\frac{dE_w}{dt} &= \int d\mathbf{r} \rho_d \mathbf{v}_d \cdot \mathbf{F}_{ext} = \int d\mathbf{r} \boldsymbol{\nabla} \cdot (\rho_d \mathbf{v}_d) U_h^1 \\
&= -\int d\mathbf{r} \frac{\partial \rho_d}{\partial t} U_h^1 = G \int d\mathbf{r} d\mathbf{r}' \frac{\partial \rho_d}{\partial t} \frac{\rho_h^1}{|\mathbf{r} - \mathbf{r}'|} \\
&= -\int d\mathbf{r} \rho_h^1 \frac{\partial U_d}{\partial t} = -\int d\mathbf{r} \rho_h^1 \frac{\partial U_w}{\partial t}.
\end{aligned}
\tag{49}
$$

where we have integrated by parts and used the continuity equation for the disk mass; the subscript 'd' denotes disk, and in the last line we have replaced the disk potential $U_d$ by the warp potential $U_w$ since the potential of the unwarped disk is stationary. To first order in the perturbation strength we may write $\rho_h^1 = -\boldsymbol{\nabla} \cdot (\rho_h^0 \Delta \mathbf{r})$, where $\Delta \mathbf{r}(\mathbf{r}, t)$ is the Lagrangian displacement of the halo stars caused by the perturbation. The last line of equation (49) may now be written

$$
\frac{dE_w}{dt} = -\int d\mathbf{r} \rho_h^0 \Delta \mathbf{r} \cdot \boldsymbol{\nabla} \frac{\partial U_w}{\partial t}.
\tag{50}
$$

There are two equivalent ways to determine the dynamical friction exerted by the halo on the disk: (i) calculate the density and potential perturbation induced in the halo by the disk and then integrate the resulting gravitational force from the halo over the disk mass distribution—this is analogous to the standard discussion of Landau damping in a collisionless plasma (*cf.* Ichimaru 1973) and corresponds to the first line in equation (49); (ii) calculate the rate of change of energy of each halo star due to the warp, and integrate these energy changes over the halo (which corresponds to eq. 50). We use the second approach in this paper.

Our calculations are based on the action-angle formalism developed in papers by Lynden-Bell & Kalnajs (1972), Tremaine & Weinberg (1984), and Weinberg (1986), which may be consulted for details. Our analysis employs three (not very good) approximations: (i) the warp is small, so that we can use perturbation theory in which the small parameter is the amplitude of the warp; in this approximation both $\Delta \mathbf{r}$ and $\partial U_d/\partial t$ are proportional to the warp amplitude so $dE_w/dt$ is second-order in the amplitude (this is the same order as the energy of the warp computed in Section 2.2; thus the damping rate $\dot{E}_w/E_w$ is independent of amplitude); (ii) the unperturbed potential (halo plus flat disk) is spherical; this is certainly incorrect—and moreover inconsistent, since we assume that the halo is axisymmetric but non-spherical when estimating the pattern speed and energy of bending modes—but orbits in spherical potentials are easier to analyze than orbits in flattened potentials and the dynamical friction force should not depend strongly on the shape of the halo orbits; (iii) we neglect the self-gravity of the halo response; again, this is certainly incorrect but unlikely to cause more than a factor of two or so error in the results.

In the unperturbed state (no warp) the motion of the halo stars is described by an autonomous Hamiltonian $H_0$. The phase-space coordinates of a star are specified using the action-angle variables of this Hamiltonian $(\mathbf{I}, \mathbf{w})$; in these coordinates $H_0 = H_0(\mathbf{I})$ so that Hamilton's equations become

$$
\dot{\mathbf{I}} = -\frac{\partial H_0}{\partial \mathbf{w}} = 0, \qquad \dot{\mathbf{w}} = \frac{\partial H_0}{\partial \mathbf{I}} \equiv \boldsymbol{\Omega}(\mathbf{I}),
\tag{51}
$$

with solutions

$$\mathbf{I}(t) \equiv \mathbf{I_0} = \text{constant} \qquad \mathbf{w}(t) \equiv \mathbf{w_0}(t) = \mathbf{\Omega}(\mathbf{I_0})t + \text{constant.} \tag{52}$$

The density of halo stars in phase space is described by a distribution function (hereafter DF) $F$ and Jeans' theorem implies that $F = F(\mathbf{I})$.

When a warp is present the Hamiltonian becomes

$$H = H_0 + U_w. \tag{53}$$

Hamilton's equations become

$$\dot{\mathbf{I}} = -\frac{\partial U_w}{\partial \mathbf{w}}, \qquad \dot{\mathbf{w}} = \mathbf{\Omega}(\mathbf{I}) + \frac{\partial U_w}{\partial \mathbf{I}}. \tag{54}$$

The perturbing potential can be expanded in an action-angle Fourier series (*cf.* Appendix A),

$$U_w(\mathbf{I}, \mathbf{w}, t) = \text{Re} \left\{ \sum_{l_3=0}^{\infty} \sum_{l_1, l_2=-\infty}^{\infty} \Psi_{l_1 l_2 l_3}(\mathbf{I}) e^{i(\mathbf{l} \cdot \mathbf{w} - \omega_{\mathbf{l}} t)} \right\}. \tag{55}$$

The first-order corrections to the orbit (52) caused by the perturbation are

$$\Delta \mathbf{I} = -\frac{\partial \chi}{\partial \mathbf{w}}, \qquad \Delta \mathbf{w} = \frac{\partial \chi}{\partial \mathbf{I}} \tag{56}$$

where

$$\chi = \int_{-\infty}^{t} U_w dt = \text{Re} \left\{ \sum_{l_3=0}^{\infty} \sum_{l_1, l_2=-\infty}^{\infty} \Psi_{l_1 l_2 l_3} \frac{e^{i(\mathbf{l} \cdot \mathbf{w} - \omega_{\mathbf{l}} t)}}{i(\mathbf{l} \cdot \mathbf{\Omega} - \omega_{\mathbf{l}})} \right\}, \tag{57}$$

and these quantities are evaluated along the unperturbed orbit $(\mathbf{I_0}, \mathbf{w_0})$. We have assumed that $\omega_{\mathbf{l}}$ has a small positive imaginary part, $\eta$, as if the potential were turned on slowly in the distant past.

The integrand of equation (50) may be written

$$\Delta \mathbf{r} \cdot \mathbf{\nabla} \frac{\partial U_w}{\partial t} = \frac{\partial^2 U_w}{\partial \mathbf{I} \partial t}(\mathbf{I_0}, \mathbf{w_0}, t) \Delta \mathbf{I} + \frac{\partial^2 U_w}{\partial \mathbf{w} \partial t}(\mathbf{I_0}, \mathbf{w_0}, t) \Delta \mathbf{w}. \tag{58}$$

We evaluate this expression using equations (55)–(57). Since the DF depends only on the actions we can average the result over the unperturbed angles $\mathbf{w_0}$. As described in Tremaine & Weinberg (1984), the terms in this expression involve combinations of periodic exponentials, most of which vanish upon averaging over initial phases[5]. Combining the surviving terms we find

$$\left\langle \Delta \mathbf{r} \cdot \mathbf{\nabla} \frac{\partial U_w}{\partial t} \right\rangle_{\mathbf{w}} = \sum_{l_3=0}^{\infty} \sum_{l_1, l_2=-\infty}^{\infty} \mathbf{l} \cdot \frac{\partial}{\partial \mathbf{I}} \left\{ |\Psi_{\mathbf{l}}(\mathbf{I})|^2 \frac{2(\omega_{\mathbf{l}} - \mathbf{l} \cdot \mathbf{\Omega})\eta e^{2\eta t}}{2(|\mathbf{l} \cdot \mathbf{\Omega} - \omega_{\mathbf{l}}|^2 + \eta^2)} \right\}, \tag{59}$$

---

[5] Or are oscillatory in time, in the case $l_3 = 0$.

where $\omega_{\mathbf{l}}$ is now taken to be real as the contribution from the small imaginary part $\eta$ is written out explicitly. We now use the identity

$$\lim_{\eta \to 0^+} \frac{\eta}{x^2 + \eta^2} = \pi \delta(x). \tag{60}$$

Integrating over the DF of halo stars finally gives the rate of change of the disk energy[6] (cf. eq. 50),

$$
\begin{aligned}
\dot{E}_d &= -\int d\mathbf{I} d\mathbf{w} F(\mathbf{I}) \Delta \mathbf{r} \cdot \nabla \frac{\partial U_w}{\partial t} = -(2\pi)^3 \int d\mathbf{I} F(\mathbf{I}) \left\langle \Delta \mathbf{r} \cdot \nabla \frac{\partial U_w}{\partial t} \right\rangle_{\mathbf{w}} \\
&= 4\pi^4 \sum_{l_3=0}^{\infty} \sum_{l_1, l_2 = -\infty}^{\infty} \int d\mathbf{I} (\mathbf{l} \cdot \mathbf{\Omega}) \mathbf{l} \cdot \frac{\partial F(\mathbf{I})}{\partial \mathbf{I}} |\Psi_{\mathbf{l}}(\mathbf{I})|^2 \delta(\mathbf{l} \cdot \mathbf{\Omega} - \omega_{\mathbf{l}}) \\
&\quad - 4\pi^4 \sum_{l_3=0}^{\infty} \sum_{l_1, l_2 = -\infty}^{\infty} \int d\mathbf{S} \cdot \mathbf{l} (\mathbf{l} \cdot \mathbf{\Omega}) F(\mathbf{I}) |\Psi_{\mathbf{l}}(\mathbf{I})|^2 \delta(\mathbf{l} \cdot \mathbf{\Omega} - \omega_{\mathbf{l}}),
\end{aligned} \tag{61}
$$

where $d\mathbf{S}$ is the area element on the boundary of action space and we have set $t = 0$. For most DFs the last (surface) term vanishes, as the following argument shows: Let $d\mathbf{S}$ be an area element on the boundary of action space. The flux of stars through this element with angles in the interval $d\mathbf{w}$ is $d\mathbf{w} d\mathbf{S} \cdot \dot{\mathbf{I}} F(\mathbf{I})$; using equations (54) and (55) this flux equals $-\sum_{\mathbf{l}} d\mathbf{w} d\mathbf{S} \cdot \mathbf{l} F(\mathbf{I}) \mathrm{Re} \{ i \Psi_{\mathbf{l}}(\mathbf{I}) \exp[i(\mathbf{l} \cdot \mathbf{w} - \omega_{\mathbf{l}} t)] \}$. The terms in this sum are linearly independent functions of $\mathbf{w}$. Since stars cannot leave action space, each term in the sum must be zero. Thus $d\mathbf{S} \cdot \mathbf{l} F(\mathbf{I}) \Psi_{\mathbf{l}}(\mathbf{I})$ must vanish, which implies that the surface term in (61) must vanish. (This argument can fail if the DF diverges on the boundary of action space.) If the surface term vanishes, equation (61) simplifies to

$$\dot{E}_d = 4\pi^4 \sum_{l_3=0}^{\infty} \sum_{l_1, l_2 = -\infty}^{\infty} \int d\mathbf{I} (\mathbf{l} \cdot \mathbf{\Omega}) \mathbf{l} \cdot \frac{\partial F(\mathbf{I})}{\partial \mathbf{I}} |\Psi_{\mathbf{l}}(\mathbf{I})|^2 \delta(\mathbf{l} \cdot \mathbf{\Omega} - \omega_{\mathbf{l}}). \tag{62}$$

Only resonant orbits with $\mathbf{l} \cdot \mathbf{\Omega}(\mathbf{I}) = \omega_{\mathbf{l}}$ contribute to the energy transfer; a particular resonant triplet $\mathbf{l} = (l_1, l_2, l_3)$ contributes to a net damping or growth of the perturbation depending on the components of $\mathbf{\Omega}(\mathbf{I})$ and the gradient of $F(\mathbf{I})$ along the resonant triplet.

Finally, we note that the formalism we have developed is related to Goodman's (1988) stability criterion for galaxies, which is based on the sign of a quadratic functional of a trial perturbation. Consider a trial perturbation $U_w$ with time dependence $e^{\eta t}$ with $\eta$ real (not $e^{-im\Omega_p t + \eta t}$ as we have assumed so far). Then Goodman shows that a sufficient criterion for instability is

$$\dot{E}_w > \frac{\eta G}{2} \int d\mathbf{R} d\mathbf{R}' \mu(R) \mu(R') \frac{[Z(\mathbf{R}, t) - Z(\mathbf{R}', t)]^2}{|\mathbf{R} - \mathbf{R}'|^3}, \tag{63}$$

---

[6] Our result is consistent with equation (65) of Tremaine & Weinberg (1984), who assume $l_3 \neq 0$ and a single pattern speed, so that $\omega_{\mathbf{l}} = l_3 \Omega_p$ and the torque is related to the rate of energy change by $\dot{E} = \Omega_p \tau_z$ (Jacobi's integral). We have assumed that the evolution of the warp is sufficiently rapid that we are in the 'fast' limit of Tremaine and Weinberg, that is, that the damping rate is large compared to the perturbation strength; this assumption is always justified if the warp amplitude to sufficiently small, because the damping time is independent of amplitude.

where $\dot{E}_w$ is defined by equation (49) and evaluated using equation (59)—compare Goodman's equation (24).

### 3.1    Halos with isotropic distribution functions

An important special case is an unperturbed halo DF that depends on energy alone, $F = F(E) = F[H_0(\mathbf{I})]$. Then equations (51) and (62) imply

$$\dot{E}_d = 4\pi^4 \sum_{l_3=0}^{\infty} \sum_{l_1, l_2 = -\infty}^{\infty} \int d\mathbf{I}(\mathbf{l} \cdot \boldsymbol{\Omega})^2 \frac{dF}{dE} |\Psi_\mathbf{l}(\mathbf{I})|^2 \delta(\mathbf{l} \cdot \boldsymbol{\Omega} - \omega_\mathbf{l}). \tag{64}$$

In most model galaxies with $F = F(E)$ the halo DF is a decreasing function of energy, $dF/dE < 0$; following Goodman (1988), we call these 'IDDF halos' (for 'isotropic, decreasing distribution function'). In IDDF halos, equation (64) implies that $\dot{E}_d < 0$. In other words, *in IDDF halos, a warped disk always loses energy to the halo.* Thus disk disturbances with positive energy damp, while disturbances with negative energy grow.

Equation (22) shows that bending disturbances with a single azimuthal wavenumber $m \geq 0$ have positive energy if $\nu_h(R) > m\Omega(R)$ at all radii. Thus in IDDF halos, dynamical friction damps all bending disturbances so long as

$$\nu_h(R) > m\Omega(R) \quad \text{at all } R. \tag{65}$$

For $m = 1$ disturbances, equation (25) implies that the damping criterion (65) can be replaced by the stronger damping criterion

$$\nu_h(R) > \Omega_h(R) \quad \text{at all } R, \tag{66}$$

which is the condition that the halo is vertically stiff. In other words, *in vertically stiff IDDF halos, all $m = 1$ bending disturbances damp.*

Bertin & Mark (1980) examined the evolution of short-wavelength bending waves of a thin disk embedded in a halo. The halo was assumed to have an isotropic Maxwellian velocity distribution, and the disk dynamics were analyzed in the WKB approximation. They concluded that dynamical friction from the halo would excite all bending waves propagating inside the corotation radius. Toomre (1983) has pointed out that the validity of this conclusion is suspect, because Bertin and Mark ignored the halo's contribution to the vertical frequency, effectively setting $\nu_h = 0$ and thereby ensuring that the halo was vertically soft. The more general approach developed in this paper allows a rapid derivation of the correct answer: if, like Bertin and Mark, we assume an IDDF halo, then bending waves are excited by dynamical friction from the halo if and only if they have negative energy. From equation (46) the energy density of a WKB bending wave is negative if and only if

$$0 < \Omega_p < \Omega(R), \tag{67}$$

i.e. for prograde waves inside corotation.

Equation (67) is the condition for excitation of WKB bending waves for any IDDF halo. The WKB dispersion relation (43) for $m = 1$ waves implies that $|\Omega - \Omega_p| \geq \nu_h$, so a necessary condition for excitation is $\nu_h < \Omega$, which is consistent with the damping criterion (65)[7]. These results imply that the excitation of WKB bending waves by the halo is considerably more difficult than Bertin & Mark (1980) assumed.

We shall not pursue the analysis of WKB bending waves further, since the WKB approximation is invalid for most observed galactic warps. Instead we shall focus on numerical evaluation of damping and excitation rates for more realistic warps.

## 3.2 A model for the halo

In this section we describe the analytic model for the halo DF that we shall use in our numerical estimates of damping rates.

In a universe dominated by collisionless dark matter, galaxy halos form by the gravitational collapse of random peaks in the cosmological density field (cf. Warren et al. 1992 and references therein). Both analytic arguments and numerical simulations show that halos have power-law density profiles that are roughly consistent with the profiles $\rho \propto r^{-2}$ implied by observations of flat rotation curves of spiral galaxies. The simulated halos tend to be triaxial with minor to major axis ratios $c/a \sim 0.5$, although the uncertain effects of dissipation are likely to reduce the triaxiality (Katz & Gunn 1991; Dubinski 1994). The velocity ellipsoid in the halo is radially elongated: the ratio of the mean-square radial velocity to the mean-square tangential velocity is given by $2\langle v_r^2 \rangle / \langle v_t^2 \rangle \simeq 1.4$ (Dubinski 1992), whereas an isotropic dispersion tensor would have the value unity. Tidal torquing from neighboring structures gives the halo a net angular momentum; Dubinski finds that the ratio of the maximum rotation velocity to the central line-of-sight velocity dispersion is roughly correlated with the dimensionless spin parameter, $v_{\max}/\sigma_c \approx 4\lambda$, where $\lambda = J|E|^{1/2}G^{-1}M^{-5/2}$ has a median value of about 0.05 (White 1984; Barnes & Efstathiou 1987; Zurek et al. 1988).

To simplify our calculations we ignore the intrinsic triaxiality of the halo and the contribution of the disk to the total potential: thus we assume a halo DF, halo orbits, and a rotation curve for the disk that are consistent with a spherical potential provided entirely by the halo. This approximation is not strictly consistent, since we assume that the halo is axisymmetric but non-spherical when estimating the pattern speed of bending modes or calculating the mode energy (see eq. 75 at the end of this subsection). The bias caused by the spherical approximation should lead us to *under*estimate the damping and excitation rates in the likely case that the halo is flattened towards the disk plane, since the halo density will be higher and the velocity of the halo stars will be lower than we have assumed.

---

[7]In the WKB approximation the damping criteria (65) and (66) are generally the same when $m = 1$, because there is no distinction between $\Omega_h$ and $\Omega$. The reason is that the dispersion relation (43) can usually be satisfied in the WKB limit ($|KR| \gg 1$) only if the surface density $\mu$ is small, so the disk's contribution to the total rotation speed $\Omega$ is negligible.

Our model for the halo DF: (i) assumes that the halo potential generates a flat rotation curve with circular speed $v_c$,

$$U_h(r) = v_c^2 \ln\left(\frac{r}{r_s}\right), \tag{68}$$

which implies a halo density

$$\rho(r) = \frac{\nabla^2 U(r)}{4\pi G} = \frac{v_c^2}{4\pi G r^2} \tag{69}$$

($r_s$ is a fiducial radius, which eventually scales out of the problem); (ii) allows for a range of velocity anisotropy, expressed by the ratio $2\langle v_r^2\rangle/\langle v_t^2\rangle$; (iii) allows a non-zero mean rotation speed $\langle v_\phi\rangle$, which is taken to be independent of radius.

The DF that we use is

$$F(E, J, J_z) = F_0 J^{2(\gamma-1)} e^{-E/\sigma^2}(1 + \alpha J_z/J), \tag{70}$$

where $E$ is the energy, $J$ is the angular momentum, and $J_z$ is the component of angular momentum normal to the unperturbed disk. The parameter $\alpha$ controls the halo rotation, while $\gamma$ controls the velocity anisotropy; we must have $|\alpha| \leq 1$ so that the DF is non-negative, and $\gamma > 0$ so that the spatial density is finite[8]. In order that this DF generates the density (69) we must have

$$\sigma^2 = \frac{v_c^2}{2\gamma}, \qquad F_0 = \frac{1}{4\pi^{5/2}r_s^2 G v_c} \frac{\gamma^{\gamma+1/2}}{J_s^{2(\gamma-1)}\Gamma(\gamma)}, \tag{71}$$

where $J_s = r_s v_c$. In the special case $\alpha = 0$, $\gamma = 1$, the DF and potential reduce to those of the singular isothermal sphere (the normalization is the same as Weinberg 1986, eq. 24). In turn, $F$ is a special case of the larger classes of scale-free DFs discussed by Gerhard (1991) and Evans (1994).

The mean-square velocities in the radial and tangential directions are

$$\langle v_r^2\rangle = \sigma^2, \qquad \langle v_t^2\rangle = 2\gamma\sigma^2, \tag{72}$$

so the ratio of the principal axes of the velocity ellipsoid is given by

$$\frac{2\langle v_r^2\rangle}{\langle v_t^2\rangle} = \frac{1}{\gamma}; \tag{73}$$

---

[8]Galaxies with predominantly radial orbits are likely to be unstable; in fact Palmer & Papaloizou (1987) have shown that all galaxies with distribution functions that are unbounded as $J \to 0$ are unstable. However their proof assumes that the resonant combination of frequencies $\Omega_1 - 2\Omega_2 \propto J$ as $J \to 0$; while this is generally true for galaxies with a flat central core, in the singular isothermal sphere $\Omega_1 - 2\Omega_2 \propto 1/|\ln J|$ as $J \to 0$, and the Palmer-Papaloizou proof does not apply. In any case, it is not crucial for our purposes to work with stable halo models, because the self-gravity of the halo response—which is central to the instability—is neglected in our calculations of dynamical friction.

a halo with nearly radial orbits corresponds to $\gamma \to 0$ while one with nearly circular orbits has $\gamma \to \infty$. The mean rotation speed is

$$\frac{\langle v_\phi \rangle}{v_c} = \frac{\alpha}{2\gamma^{1/2}} \frac{\Gamma(\gamma + \frac{1}{2})}{\Gamma(\gamma)} \sin\theta. \tag{74}$$

The cosmological simulations described above suggest that typical parameter values are $\gamma \simeq 0.7$ (from the velocity anisotropy) and $\alpha \simeq 0.4$ (we have identified $v_{max}$ with $\langle v_\phi \rangle (\theta = \frac{1}{2}\pi)$ and $3\sigma_c^2$ with $\langle v_r^2 \rangle + \langle v_t^2 \rangle$).

Finally, when estimating the pattern speed or energy of a bending mode, we generalize the expression for the halo potential (68) to an axisymmetric halo:

$$U_h(R, z) = \frac{1}{2}v_c^2 \ln\left(R^2 + z^2/q_\Phi^2\right), \tag{75}$$

where $q_\Phi$ is the axial ratio of the equipotential surfaces and $\epsilon_\Phi \equiv 1 - q_\Phi$ is their ellipticity ($\epsilon_\Phi > 0$ for oblate halos and $< 0$ for prolate halos). In this case the vertical and azimuthal frequencies in the field of the halo are (eqs. 2 and 5)

$$\nu_h = \frac{v_c}{q_\Phi R}, \qquad \Omega_h = \frac{v_c}{R}. \tag{76}$$

### 3.3   A model for the warp

To calculate the damping rate in a realistic warped galaxy we need to know the shape of the warp $h(R)$ (eq. 7) and its associated gravitational potential $U_w(\mathbf{r})$. Rather than solving equation (8) to obtain an exact eigenfunction, we have chosen a simple analytic parametrization of the warp. We assume that the surface density of the disk is exponential,

$$\mu(R) = \mu_0 e^{-R/R_d}, \tag{77}$$

so that the total disk mass $M_d = 2\pi\mu_0 R_d^2$. We set the parameter

$$\frac{GM_d}{R_d v_c^2} = 1.6, \tag{78}$$

in all of our simulations, which is roughly correct for our Galaxy. We write $h(R) = R_d f(R/R_d)$ where $R_d$ is the scale length of the disk. If the mode has real eigenfrequency then we can assume that $h(R)$ is real (by the anti-spiral theorem of Lynden-Bell & Ostriker 1967); in this case the line of nodes is radial, which is approximately consistent with observations.

Our parametrization has the form

$$f(x) = Cx^n e^{-\lambda x}; \tag{79}$$

where $C$ is an arbitrary normalization (recall that the damping rate is independent of the amplitude). This parametrization permits us to model two distinct types of warps. (i) For $n = 1$ the inner part of the disk is tilted relative to the halo (as in the Laplace mode) and the outer disk is flat; in this case equation (79) gives a warp whose curvature in its outer parts is more gradual than observed warps (see Figure 1a), but this is not a major defect since most of the interaction with the halo comes from the tilted inner disk, which is not curved either in our model or in real galaxies. (ii) For $n \gg 1$ the inner disk is flat and the outer disk is tilted. In this case the shape of the warp can be made to closely resemble observed warps (except that at very large radii the disk becomes flat again—its maximum height is at $x_{max} = n/\lambda$—but if $x_{max} \gg 1$ the surface density in the region where the disk becomes flat again will be negligible).

An advantage of the functional form (79) is that the spherical harmonic expansion of the warp potential $U_{ln}(r)Y_{ln}(\theta, \phi)$ can be evaluated in terms of incomplete gamma functions (see Appendix B).

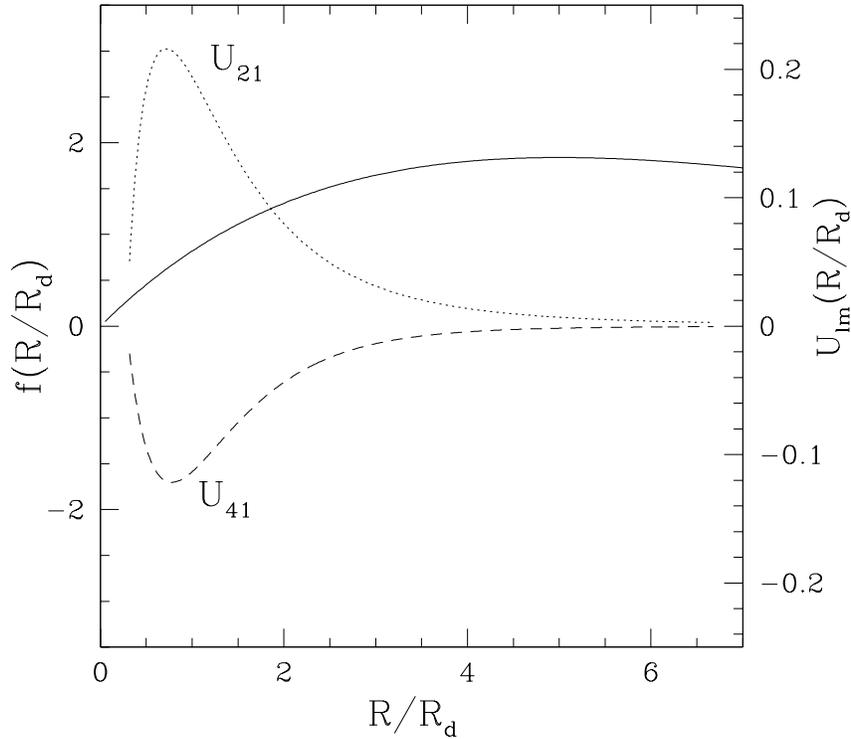

Fig. 1a.— The solid line shows the shape of a typical warp used in our estimates of the warp damping/excitation rate, derived from eq. (79) with $C = 1$, $n = 1$, $\lambda = 0.2$. The dotted and dashed lines indicate the first two non-zero components, $U_{21}(r), U_{41}(r)$, of the spherical harmonic expansion of the potential of the warped disk with $m = 1$; the scale for the potentials is shown on the right hand axis in units of $GM_d/R_d$.

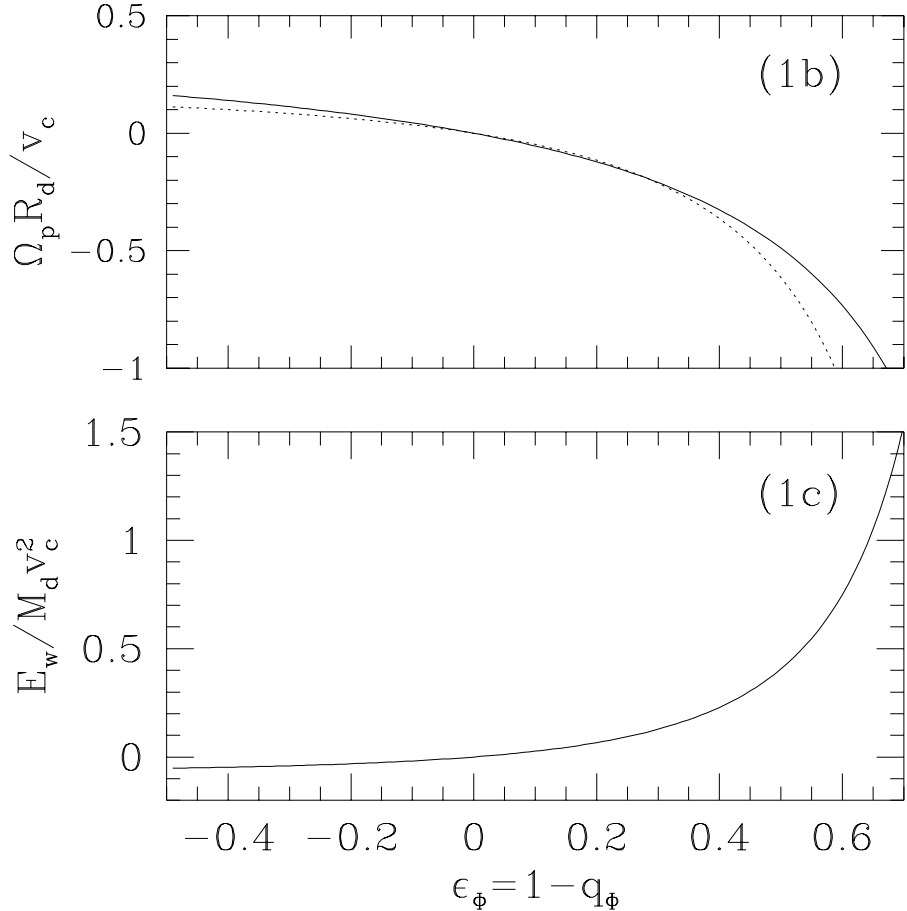

Fig. 1b.— The pattern speed in units of $v_c/R_d$ (the angular speed of the disk at one scale length), as a function of the ellipticity of the halo potential. The pattern speed is computed using the warp shape in Figure 1a and eq. (14)—the solution of the quadratic equation with the smaller value of $|\Omega_p|$ is shown. The dotted line corresponds to the Sparke-Casertano formula (eq. 17).

Fig. 1c.— Energy of the warp shown in (a) as a function of the ellipticity of the halo potential, computed from eqs. (29), (77) and (78).

Figure 1a shows an example of a warp shape parametrized by equation (79) with $n = 1$ and $\lambda = 0.2$, along with the radial potential functions associated with the first two nonvanishing $m = 1$ spherical harmonics, $U_{21}(r)$, $U_{41}(r)$. Figure 1b shows the pattern speed for this warp, calculated using equation (14), as a function of the halo ellipticity $\epsilon_\Phi$ (eq. 75). The dotted line shows the Sparke-Casertano formula for the pattern speed (eq. 17), which is a good approximation to the more accurate formula. Figure 1c shows the energy of the warp (eq. 29) with $C = 1$. The figure illustrates that warps embedded in vertically stiff halos ($\epsilon_\Phi > 0$) generally precess in a retrograde direction ($\Omega_p < 0$) and have positive energy, while warps embedded in vertically soft halos ($\epsilon_\Phi < 0$) precess in a prograde direction ($\Omega_p > 0$) and have negative energy.

## 4 Calculation of Damping and Excitation Rates

We now calculate the rate of change of the warp energy, and the resulting damping or growth rate caused by dynamical friction from the halo. Weinberg (1986) has described similar calculations, based on a formula analogous to (62), for the rate of decay of a satellite on a circular orbit in a spherical potential. Thus we have simply summarized the principal formulae in Appendix A and refer the reader to Weinberg (1986) for details.

The formula (62) for the rate of change of the warp energy is combined with equation (29) for the energy to obtain the characteristic evolution rate

$$\Gamma = \frac{1}{E_w}\frac{dE_w}{dt};\tag{80}$$

$\Gamma > 0$ implies excitation and $\Gamma < 0$ implies damping. In our model the evolution rate $\Gamma$ is a function of the velocity anisotropy of the halo (parametrized by $\gamma$), the rotation of the halo (parametrized by $\alpha$), the warp pattern speed (written in dimensionless form as $\Omega_p R_d/v_c$), and the shape of the bending disturbance; it is proportional to the mass of the disk but independent of the amplitude of the disturbance (for small perturbations) since both the energy and the rate of energy change scale with the square of the amplitude.

We evaluate the action-angle transform of the perturbing potential using equations (A1)–(A3) and (A23), along with equations (79) and (B9)–(B10) for the warp shape and potential. We treat the integrals in equations (A10), (A11) and (A23) as the solutions to a coupled set of ordinary differential equations, which we integrate using an adaptive Runge-Kutta routine. We use Romberg integration to evaluate the integral in equation (A26), summing over the possible resonant triplets in the integrand itself, rather than evaluating an integral for each term separately. Since the number of resonant triplets grows $\propto l^3$, we must cut off the summation at a relatively small maximum harmonic $l_{\max}$. Experiments on a few points show that $\dot{\mathcal{E}}(l_{\max}) = \dot{\mathcal{E}} + Cl_{\max}^{-\beta}$, with $\beta \simeq 1$. In practice we evaluated $\dot{\mathcal{E}}(l_{\max})$ at $l_{\max} = 2, 4$, and 6, fitting these values to a quadratic in $l_{max}^{-1}$, and extrapolate to $l_{\max} \to \infty$. We believe that the damping/excitation rates we have obtained are accurate to within $\pm 3\%$.

We have verified our results in the epicycle limit $(1 - \kappa \ll 1)$ where the functional form for the potential transforms becomes analytic, and also compared our formulas with similar transforms calculated by Weinberg (1985).

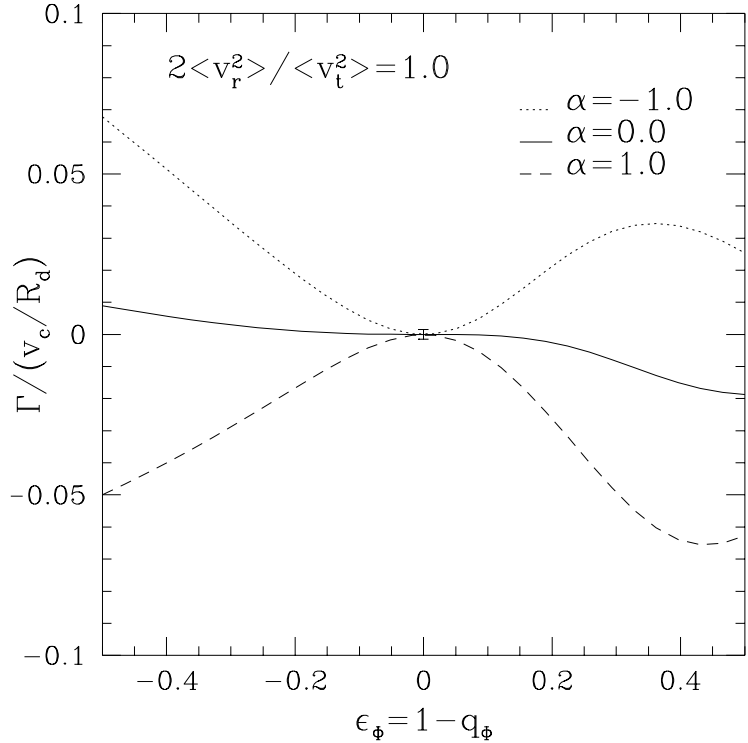

Fig. 2.— The damping rate $\Gamma = \dot{E}_w / E_w$ as a function of the ellipticity of the halo potential. The warp shape is shown in Figure 1a. The halo DF is described by eq. (70) with $\gamma = 1$ (isotropic velocity distribution). Warps with $\Gamma > 0$ grow, while those with $\Gamma < 0$ damp. Damping or excitation occurs within a Hubble time if $|\Gamma| > 0.0015$ (assuming $R_d / v_c = 1.5 \times 10^7$ y). The small errorbar centered at $\epsilon_\Phi = 0$ corresponds to the range of damping rates where damping or excitation does not occur within a Hubble time. The dotted, solid, and dashed lines correspond respectively to halos with rotation parameters $\alpha = -1, 0, 1$ described in the text; these values span the range of possible rotation rates.

Figure 2 shows the dimensionless damping/excitation rate $\Gamma R_d / v_c$ as a function of the halo ellipticity $\epsilon_\Phi$. The warp has the shape shown in Figure 1a and is embedded in a halo with an isotropic velocity distribution, $2\langle v_r^2 \rangle / \langle v_t^2 \rangle = 1$. For parameters corresponding to those of our Galaxy, $R_d / v_c = 1.5 \times 10^7$ y, the damping/excitation time is less than $10^{10}$ y when $|\Gamma| R_d / v_c > 0.0015$.

We plot results for three values of $\alpha$ which span the range of possible rotation rates consistent with our model (eq. 74). First consider the non-rotating halo ($\alpha = 0$) with isotropic velocity distribution (an IDDF halo in the notation of Section 3.1). As shown in that section, a warp always loses energy to an IDDF halo. Thus warps with positive energy damp. However, if the halo is vertically soft ($\epsilon_\Phi < 0$, which obtains if the disk lies in the equatorial plane of a prolate halo potential), the energy of the warp is negative (Figure 1c). Thus, a warp embedded in a non-rotating, vertically soft IDDF halo can grow in amplitude, while one embedded in a vertically stiff halo must damp.

Halo rotation tends to increase the rate of change of warp energy $\dot{E}_w$ if the halo rotates in the same sense as the pattern speed, and decreases $\dot{E}_w$ if the halo rotates in the opposite sense. Thus, adding rotation tends to increase (or decrease) the warp amplitude if the sign of $\alpha \Omega_p / E_w$ is positive (negative). If the pattern speed is small, $E_w / \Omega_p$ is always negative (eq. 29); thus halo rotation contributes to warp growth only if the halo rotates in the opposite sense to the disk.

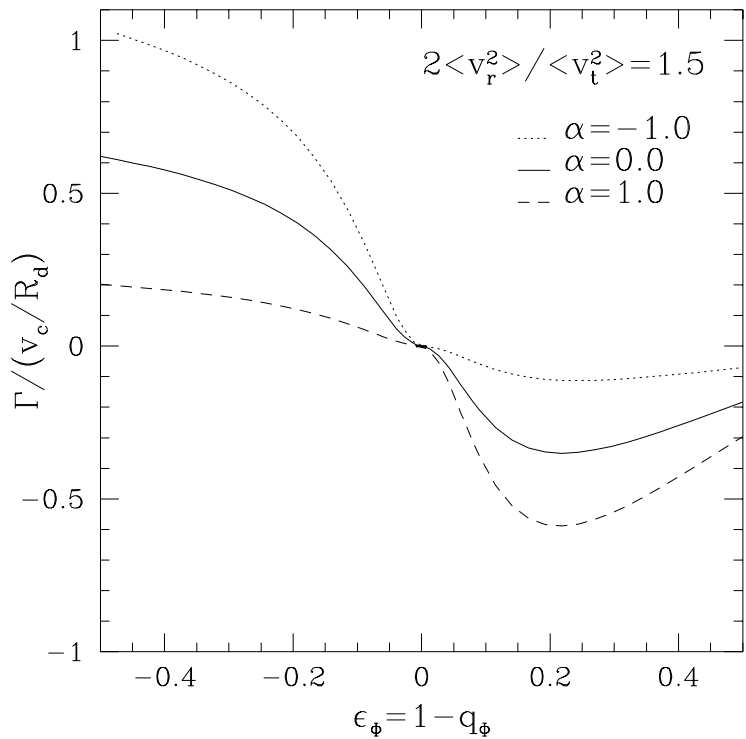

Fig. 3.— Same as Figure 2, but with for a halo with predominantly radial orbits, $2\langle v_r^2 \rangle / \langle v_t^2 \rangle = 1.5$. Note that the vertical range of the graph is a factor of ten larger than Figure 2.

Figure 3 shows the damping/excitation rates for warps embedded in halos with predominantly radial orbits, $2\langle v_r^2 \rangle / \langle v_t^2 \rangle = 1.5$. The damping rate is much larger than in the isotropic case; in fact so large ($|\Gamma| \gtrsim |\Omega_p|$) that our treatment of the warp as a slowly evolving normal mode is suspect.

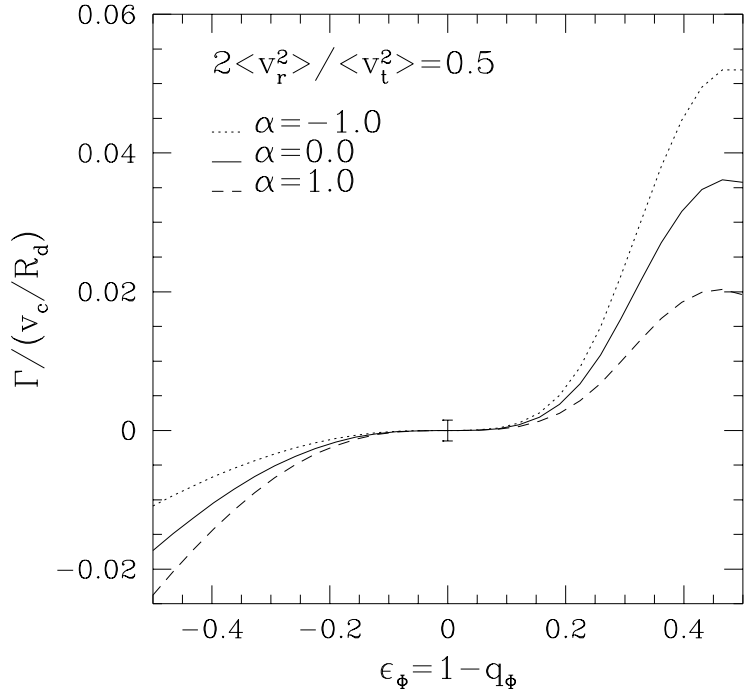

Fig. 4.— Same as Figure 2, but with for a halo with predominantly circular orbits, $2\langle v_r^2 \rangle / \langle v_t^2 \rangle = 0.5$. Note that the vertical range is only half as large as Figure 2.

Figure 4 shows the damping/excitation rates for warps embedded in halos with predominantly circular orbits, $2\langle v_r^2 \rangle / \langle v_t^2 \rangle = 0.5$. In this case energy flows from the halo to the warp, a consequence of the strong anisotropy in the phase space distribution function. Thus positive energy warps ($\epsilon_\Phi > 0$) grow, while negative energy warps damp.

Figure 5 shows the dimensionless damping rate as a function of the halo velocity anisotropy, for a warped disk with a retrograde pattern speed, $R_d \Omega_p / v_c = -0.5$; this pattern speed corresponds to a halo whose equipotentials have ellipticity $\epsilon_\Phi = 0.5$ (Figure 1b). For halos with predominantly circular orbits the warp is excited, while in halos dominated by radial orbits it is damped. An interesting feature is that the damping rate diverges as $\langle v_r^2 \rangle / \langle v_t^2 \rangle \to 2$, corresponding to $\gamma \to \frac{1}{2}$. We show in Appendix C that the apparent divergence is an artifact of the perturbation theory that we have used: according to equation (61) the rate of energy loss $\dot{E}_w$ arising from the response of the halo to a perturbation potential $\Psi$ is $O(\Psi^2)$, but in fact the response is $O(\Psi^p)$ where $p = \min(2, 2\gamma + 1)$; thus for $\gamma < \frac{1}{2}$ the energy loss rate is of lower order than $O(\Psi^2)$. Our analysis is not powerful enough to determine the damping rate for halos with $\gamma < \frac{1}{2}$; nevertheless it is clear that the damping is very strong when the orbits of the halo stars are predominantly radial.

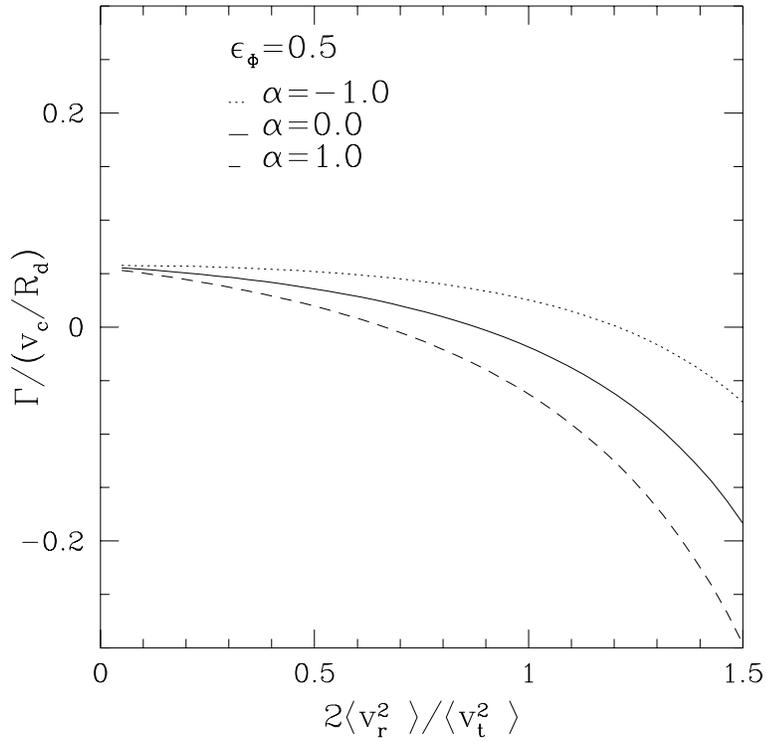

Fig. 5.— The damping rate $\Gamma$ as a function of the velocity anisotropy of the halo. The halo equipotentials are oblate, $\epsilon_\Phi = 0.5$. The dotted, solid, and dashed lines correspond to halos with dimensionless rotation $\alpha = -1.0, 0.0, 1.0$. Halos with predominantly circular orbits excite the warp, while halos with predominantly radial orbits damp the warp.

We have also examined the damping of disks that are only warped relative to the halo in their outer parts (eq. 79 with $n = 5$, $\lambda = 1$). Although the damping/excitation times are substantially longer, significant damping still generally occurs in less than a Hubble time.

## 5 Discussion and Conclusions

The most striking feature shown by Figures 2–4 is the rapid evolution of the warps from dynamical friction. All of the models we examined had damping/excitation times less than a Hubble time when the halo ellipticity was larger than 0.2, and in general the damping/excitation times were much shorter than a Hubble time. For example, in a non-rotating halo with a plausible value of the velocity anisotropy, $2\langle v_r^2 \rangle / \langle v_t^2 \rangle = 1.5$ (recall from Section 3.2 that cosmological simulations yield a typical value of 1.4 for this parameter), the damping time $-\Gamma^{-1} = 10^8$ y even when the ellipticity of the halo potential is as small as $\epsilon_\Phi = 0.2$.

Thus our numerical results show that dynamical friction between a warped galaxy disk and its surrounding dark matter halo usually leads to significant evolution of the warp over a Hubble time. The rate of evolution depends strongly on the velocity anisotropy and net rotation of the halo stars and the ellipticity of the halo potential.

Dynamical friction can either damp or excite a warp: warps can have either positive or negative energy, and friction can either add energy to or remove energy from the warp. If the disk lies in the equatorial plane of an axisymmetric halo, then the energy of an $m = 1$ warp is generally positive if the halo is oblate or vertically stiff, and negative if the halo is prolate or vertically soft. A non-rotating IDDF halo always removes energy from the warp. Halos with predominantly radial orbits remove energy more rapidly than halos with an isotropic velocity distribution, while in the unlikely case that the halo orbits are predominantly circular, dynamical friction can add energy to the warp. The effect of halo rotation is to add energy to the warp if the mean rotation rate is in the same direction as the precession of the warp, and to remove energy otherwise.

Our assumption that the halo is axisymmetric is a major oversimplification, since cosmological simulations of galaxy formation typically produce halos that have highly triaxial shapes. The formation of gaseous disks in triaxial potentials has been investigated by analytic arguments and numerical simulations (Steiman-Cameron & Durisen 1984, Habe & Ikeuchi 1985, Pearce & Thomas 1991, Thomas et al. 1994). Disks can form in the principal plane normal to the major or the minor axis (orbits in the principal plane normal to the intermediate axis are unstable). As a first approximation, we may treat the halo as vertically stiff if the disk is normal to the minor axis and vertically soft if the disk is normal to the major axis. The principal plane in which accreted gas will settle depends on its initial angular momentum; simple analytic arguments based on perturbation theory imply that the probability that gas with randomly directed angular momentum will settle into the principal plane normal to the major axis is $2I_{crit}/\pi$, where

$$\sin^2 I_{crit} = \frac{b^{-2} - a^{-2}}{c^{-2} - a^{-2}}. \tag{81}$$

Taking typical halo axis ratios of $c/a = 0.45$, $b/a = 0.7$ (Dubinski & Carlberg 1991, Warren et al. 1992), we find that the probability that disks lie in the principal plane normal to the major axis is roughly 0.3; this is an upper limit since in practice the angular momentum vector of the infalling gas is strongly correlated with that of the halo material (e.g. Katz & Gunn 1991) For the minority of galaxies in which the disk forms in the principal plane normal to the halo major axis, our results show that the disk is likely to be strongly unstable to warping; the outcome of this instability is presumably the settling of the disk into the principal plane normal to the minor axis[9]. It is possible that some warped galaxies are a manifestation of this instability.

Rotating halos can also excite warps. Warps in vertically stiff halos generally have retrograde pattern speeds (eq. 17) and positive energy, so that excitation requires that the halo and the disk rotate in opposite directions; as we have mentioned, cosmological simulations show that the angular momentum vectors of halos and their embedded disks are generally similar (although polar rings [Casertano et al. 1991] and counter-rotating disks [Rubin et al. 1992] offer counter-examples). Warps can also be excited if the halo orbits are predominantly circular, but simulations of galaxy formation suggest that halo orbits are radial rather than circular. Thus excitation by either of these mechanisms is unlikely to be the major cause of warps either.

A general difficulty with explaining warps by excitation from dynamical friction is that the excitation timescales are short compared to a Hubble time, and there is no obvious nonlinear process to limit the amplitude of the warps at their observed values.

In the majority of galaxies, for which dynamical friction damps rather than excites warps, the damping time is short compared to the Hubble time. In the plausible case that halos have predominantly radial orbits, the damping time is often shorter than even the precession time $2\pi/|\Omega_p|$ (Figure 2).

These results bring into sharp focus the problem of exciting warps. Ostriker & Binney (1989) have argued that normal galaxy formation processes will not excite the Laplace mode; our calculations show that even if the mode were excited during galaxy formation, it would by now have been damped away. Thus we require an excitation mechanism that operates throughout the life of the galaxy.

One plausible mechanism is twisting of the halo by cosmic infall (Ostriker & Binney 1989). Both analytic arguments and N-body simulations show that the direction of the angular momentum vector of a halo—and hence the orientation of its principal planes—is reoriented by infalling material with each doubling of the halo age (see Binney 1992 for a review). This steadily changing orientation is communicated to the inner halo and disk by gravitational torques; the warp arises because it is necessary to transport off-axis angular momentum between the inner and outer disk. In this model the warp amplitude is determined by the present rate of twisting rather than the past history of excitation and damping.

---

[9] A corollary is that massive polar rings should be unstable.

Our results can be applied to warps with other azimuthal wavenumbers; although we have done no calculations we can make some comments. Sparke (1994) has argued that axisymmetric ($m = 0$) 'bowl-shaped' modes may be present in spiral galaxies and polar rings. The analysis of Section 2.2 shows that all $m = 0$ warps have positive energy; hence they will be damped in an IDDF halo. Symmetry arguments show that halo rotation will not tend to excite axisymmetric modes either. We expect the damping from dynamical friction to be rapid, since the frequency of the $m = 0$ mode is usually much larger than the pattern speed of the Lagrange mode. Thus it is unlikely that $m = 0$ modes are excited by dynamical friction; and if they are excited by other mechanisms, they will be rapidly damped. For higher azimuthal wavenumbers, $m \geq 2$, there is a continuum of bending modes in realistic disks, so such modes will disperse even without damping from the halo.

We close this section with a summary of the approximations we have made:

1. We have used linear perturbation theory (in the warp amplitude), which is inadequate: observed warps may have amplitudes $\gtrsim 10°$, which represents only the *difference* in inclination between the inner disk and the outer edge. In perturbation theory the tilted disk tends to align with the equatorial plane of the halo; in reality a tilted disk with mass comparable to the inner halo has much more angular momentum than the halo, so that the inner halo tends to align with the disk rather than the disk with the halo (then over a longer timescale the outer halo continues to exert significant frictional force and eventually brings all three components to a common alignment).

2. In calculating the orbits of halo stars, we have assumed that the unperturbed potential of the halo plus disk is spherical, whereas in fact the halo potential is probably triaxial and the disk provides an additional flattened component. With a more realistic potential the friction would probably be increased, since the halo would generally be flattened towards the disk plane so its interaction with the warp would be stronger. An inconsistency in our calculations is that we assume the halo is axisymmetric but non-spherical when computing the warp energy and pattern speed.

3. We have approximated the halo potential as that of a singular isothermal sphere. This approximation leads us to underestimate the damping/excitation for warps with small pattern speeds, since the absence of a core implies that fewer halo stars are close to the $(l_1, l_2) = (1, -2)$ resonance that dominates the friction at small pattern speeds.

4. We have neglected the self-gravity of the halo response in order to simplify the friction

calculations. Including the self-gravity of the halo response should generally increase the dynamical friction (for satellites, the inclusion of the response self-gravity decreases the friction [Hernquist & Weinberg 1989], but this reflects mainly the difficulty of treating the motion of the center of mass consistently without self-gravity), and therefore strengthen our conclusion that the damping times are short.

5. We have treated the disk as razor-thin, with no internal structure. The dispersion relation for bending or corrugation waves in stellar systems with slab geometry (uniform in $x$ and $y$, self-gravitating, and symmetric about the $z = 0$ plane) has been investigated by Araki (1985), Weinberg (1991), and Toomre (1966, 1994). The bending wave is damped by near-resonant stars; if the horizontal variation of the bending wave is $\propto \exp[i(\mathbf{k} \cdot \mathbf{x} - \omega t)]$, the resonant stars are those with $\mathbf{k} \cdot \mathbf{v} = n\nu(E_z) + \omega$, where $n$ is an integer and $\nu(E_z)$ is the vertical oscillation frequency as a function of vertical energy (this approximation neglects the curvature of the epicyclic orbits in a real galaxy disk, but this should not invalidate our analysis since the epicycle frequency is small compared to $\nu$). If the typical horizontal velocity dispersion is $\sigma$, the criterion for significant damping—plenty of resonant stars—is $k\sigma/\nu(0) > 1 + |\omega|/\nu(0)$. The term $|\omega|/\nu(0)$ is usually negligible for Laplace modes, which have small pattern speeds. Taking $\lambda = 2\pi/k = 10$ kpc, $\sigma = 40$km s$^{-1}$, $\nu(0) = 3 \times 10^{-15}$ s$^{-1}$ (corresponding to a local density of 0.15M$_\odot$ pc$^{-3}$), we find $k\sigma/\nu(0) = 0.3$; which is probably too small for significant damping in a typical warp.

Many of the concerns raised by these assumptions can be addressed by N-body simulations, which complement the analytic approach described here. Dubinski & Kuijken (1994) have recently examined the evolution of tilted disks (both rigid and N-body) embedded in N-body halos. They find that the disk rapidly aligns with the inner halo, and the damping times that they observe are even shorter than the ones we calculate (probably because the disk contains most of the angular momentum, so that the halo aligns with the disk rather than vice versa). Our results suggest other worthwhile N-body experiments. It would be interesting to confirm whether warps can be excited in rotating or vertically soft halos, to test our prediction that the damping is strongest in halos with predominantly radial orbits, and to investigate the fate of massive polar rings.

Our conclusions are summarized briefly in the abstract.

We acknowledge helpful discussions with James Binney, Omer Blaes, Ray Carlberg, John Dubinski, Peter Goldreich, Linda Sparke, and Alar Toomre. This research was supported by NSERC.

## A   The Dimensionless Energy Loss Rate

We wish to evaluate the rate of energy transfer to the disk given by equation (62).

An orbit in a spherical potential can be described by the canonical actions $(I_1, I_2, I_3) = (I_r, J, J_z)$, and their conjugate angles $(w_1, w_2, w_3)$; $I_r$ is the radial action, and $J$ and $J_z$ are the magnitude of the total angular momentum and its component along $\mathbf{e}_z$ which we take to be normal to the unperturbed disk. The conjugate angles vary from 0 to $2\pi$, with frequencies given by Hamilton's equations (51).

Following Tremaine & Weinberg (1984) and Weinberg (1986), we expand the perturbing potential from the warped disk in both spherical harmonics and an action-angle Fourier series (*cf.* eq. 55)[10],

$$
\begin{aligned}
U_w(\mathbf{r}, t) &= \sum_{l=1}^{\infty} \sum_{n=-l}^{l} U_{ln}(r) Y_{ln}(\theta, 0) e^{in(\phi - \Omega_p t)} \\
&= \mathrm{Re}\left\{ \sum_{l_3=0}^{\infty} \sum_{l_1, l_2 = -\infty}^{\infty} \Psi_{l_1 l_2 l_3}(\mathbf{I}) e^{i(\mathbf{l}\cdot\mathbf{w} - \omega_1 t)} \right\}
\end{aligned} \tag{A1}
$$

where $l_3 = |n|$, $\omega_1 = l_3\Omega_p$, $U_{ln}^* = (-1)^n U_{l,-n}$, and the radial functions $U_{ln}$ are derived in Appendix B. The amplitudes in the two lines of equation (A1) are related by

$$
\Psi_{l_1 l_2 l_3}(I_1, I_2, I_3) = \sum_{l=1}^{\infty} \left( \frac{2}{1 + \delta_{l_3 0}} \right) V_{l l_2 l_3}(\beta) W_{l l_2 l_3}^{l_1}(I_1, I_2), \tag{A2}
$$

where

$$
V_{l l_2 l_3}(\beta) = r_{l_2 l_3}^{l}(\beta) Y_{l l_2}(\tfrac{1}{2}\pi, 0) i^{l_3 - l_2}, \tag{A3}
$$

$$
W_{l l_2 l_3}^{l_1}(I_1, I_2) = \frac{1}{\pi} \int_0^{\pi} dw_1 \cos[l_1 w_1 - l_2(\psi - w_2)] U_{l l_3}(r); \tag{A4}
$$

here $\psi$ is the angle in the orbital plane measured from the ascending node, and explicit formulae for the angles $w_1$ and $w_2$ are given in Tremaine & Weinberg (1984) and below (eqs. A10 and A11). Here $\cos\beta = J_z/J$ is the cosine of the orbital inclination, and $r_{l_2 l_3}^{l}(\beta)$ are rotation matrices which satisfy the orthogonality condition (*cf.* Gottfried 1979, section 34.5)

$$
\int_0^{\pi} d\beta \sin\beta\, r_{mn}^{l}(\beta) r_{mn}^{l'}(\beta) = \frac{2}{2l+1}\delta_{ll'}. \tag{A5}
$$

It can be shown that $\Psi_{-1}^* = \Psi_1$.

For the logarithmic potential given in equation (68) it is convenient to define a dimensionless radius and angular momentum (Weinberg 1986)

$$
\tilde{r} = \frac{r}{r_s} e^{-E/v_c^2}, \qquad \kappa = \frac{J}{J_c(E)}, \tag{A6}
$$

---

[10]When $n = 0$ the time-dependence must be written as $e^{-i\omega t}$ rather than $e^{-in\Omega_p t}$. For simplicity, we shall not explicitly include this special case in our formulae.

where $J_c(E) = J_s e^{(E/v_c^2 - 1/2)}$ is the maximum angular momentum possible for a given energy (a circular orbit) and $J_s = r_s v_c$. In these variables the orbital frequencies can be written in the form

$$\Omega_i(E, J) \equiv e^{-E/v_c^2} \nu_i(\kappa), \tag{A7}$$

where the new frequencies are given by $\nu_i = (v_c/r_s)\tilde{\nu}_i$,

$$\tilde{\nu}_1(\kappa) = \left\{ \frac{1}{2\pi} \oint \frac{d\tilde{r}}{(-2\ln\tilde{r} - \kappa^2/e\tilde{r}^2)^{1/2}} \right\}^{-1}, \tag{A8}$$

$$\tilde{\nu}_2(\kappa) = \frac{\tilde{\nu}_1(\kappa)\kappa}{2\pi e^{1/2}} \oint \frac{d\tilde{r}}{\tilde{r}^2(-2\ln\tilde{r} - \kappa^2/e\tilde{r}^2)^{1/2}}, \tag{A9}$$

and $\nu_3 = 0$. The integrations are taken over a complete orbit with turning points determined by the roots of the denominators.[11]

The angles are defined as a function of the dimensionless radius by

$$w_1(\tilde{r}) = \tilde{\nu}_1(\kappa) \int_{\tilde{r}_p}^{\tilde{r}} \frac{d\tilde{r}}{(-2\ln\tilde{r} - \kappa^2/e\tilde{r}^2)^{1/2}}, \tag{A10}$$

$$w_2 - \psi = \frac{\nu_2(\kappa)}{\nu_1(\kappa)} w_1 - \frac{\kappa}{e^{1/2}} \int_{\tilde{r}_p}^{\tilde{r}} \frac{d\tilde{r}}{\tilde{r}^2(-2\ln\tilde{r} - \kappa^2/e\tilde{r}^2)^{1/2}}. \tag{A11}$$

We divide the directional derivative of the DF in equation (70) into two parts, even and odd in $\cos\beta$,

$$\mathbf{l} \cdot \frac{\partial F}{\partial \mathbf{I}} = \mathbf{l} \cdot \mathbf{\Omega} \left( \frac{\partial F}{\partial E} \right)_{J, J_z} + l_2 \left( \frac{\partial F}{\partial J} \right)_{E, J_z} + l_3 \left( \frac{\partial F}{\partial J_z} \right)_{E, J} = F_1(E, J) + F_2(E, J)\cos\beta, \tag{A12}$$

where

$$F_1(E, J) = F_0 J^{2(\gamma-1)} e^{-E/\sigma^2} \left\{ -\frac{\mathbf{l} \cdot \mathbf{\Omega}}{\sigma^2} + \frac{1}{J}[2(\gamma-1)l_2 + \alpha l_3] \right\}, \tag{A13}$$

and

$$F_2(E, J) = \alpha F_0 J^{2(\gamma-1)} e^{-E/\sigma^2} \left[ \frac{(2\gamma-3)}{J} l_2 - \frac{\mathbf{l} \cdot \mathbf{\Omega}}{\sigma^2} \right]. \tag{A14}$$

Here we have used $\mathbf{\Omega} = \partial E/\partial \mathbf{I}$. Using $dI_2 dI_3 = J dJ d(\cos\beta)$, the integral over $\beta$ in equation (62) due to the first term, $F_1(E, J)$, is proportional to

$$\int_0^\pi d\beta \sin\beta |\Psi_{\mathbf{l}}(\mathbf{I})|^2 = \sum_{l=1}^\infty \frac{8}{(2l+1)(1 + \delta_{l_3 0})^2} |Y_{ll_2}(\tfrac{1}{2}\pi, 0)|^2 |W_{ll_2 l_3}^{l_1}(I_1, I_2)|^2, \tag{A15}$$

---

[11] Weinberg sometimes uses dimensionless units $r_s = \sigma = 1$; in these units $v_c = 2^{1/2}$.

which follows from equations (A2) and (A5). The integral arising from the second term, $F_2(E, J) \cos \beta$, is proportional to

$$
\begin{aligned}
\int_0^\pi d\beta \sin \beta \cos \beta |\Psi_{\mathbf{l}}(\mathbf{I})|^2 &= \frac{4}{(1 + \delta_{l_3 0})^2} \sum_{l, l' = 1}^\infty W_{l l_2 l_3}^{l_1}(I_1, I_2) W_{l' l_2 l_3}^{l_1 \; *}(I_1, I_2) Y_{l l_2}(\tfrac{1}{2}\pi, 0) Y_{l' l_2}^*(\tfrac{1}{2}\pi, 0) \\
&\times \int_0^\pi d\beta \sin \beta \cos \beta \, r_{l_2 l_3}^l(\beta) r_{l_2 l_3}^{l'}(\beta).
\end{aligned}
\tag{A16}
$$

The last integral vanishes unless $l = l' \pm 1$ (*cf.* Gottfried 1979); we show in Appendix B that the potential contains only terms odd in $l + l_3$ so this integral will vanish. Consequently, using (A15) in (62) we find

$$
\dot{E}_w = \sum_{\mathbf{l}} \sum_{l=1}^\infty \frac{32 \pi^4 \omega_{\mathbf{l}}}{(2l + 1)(1 + \delta_{l_3 0})^2} |Y_{l l_2}(\tfrac{1}{2}\pi, 0)|^2 \int dI_1 J dJ |W_{l l_2 l_3}^{l_1}|^2 F_1(E, J) \delta(\mathbf{l} \cdot \boldsymbol{\Omega} - \omega_{\mathbf{l}}).
\tag{A17}
$$

The resonance constraint can be written as a delta function in energy,

$$
\delta(\mathbf{l} \cdot \boldsymbol{\Omega} - \omega_{\mathbf{l}}) = \frac{\delta[E - E_{\mathbf{l}}(\kappa)]}{|\partial \mathbf{l} \cdot \boldsymbol{\Omega} / \partial E|_\kappa} = \frac{v_c^2}{|\omega_{\mathbf{l}}|} \delta[E - E_{\mathbf{l}}(\kappa)] \Theta\left(\frac{\mathbf{l} \cdot \boldsymbol{\nu}}{\omega_{\mathbf{l}}}\right)
\tag{A18}
$$

where

$$
E_{\mathbf{l}}(\kappa) = v_c^2 \ln\left[\frac{\mathbf{l} \cdot \boldsymbol{\nu}(\kappa)}{\omega_{\mathbf{l}}}\right]
\tag{A19}
$$

is the resonant energy associated with a particular $\kappa$ and $\Theta$ is the unit step function. We replace $dI_1$ by $dE / \Omega_1(E, J)$; then using equations (71), (A6), (A7) and (A13) we find

$$
\frac{J dJ F_1(E, J)}{\Omega_1(E, J)} = -\frac{2 F_0 J_s^{2\gamma}}{v_c^2 e^\gamma} \frac{\kappa^{2(\gamma - 1)}}{\tilde{\nu}_1(\kappa)} f(\kappa) d\kappa
\tag{A20}
$$

where

$$
f(\kappa) = \gamma \kappa \mathbf{l} \cdot \tilde{\boldsymbol{\nu}}(\kappa) - e^{1/2}[(\gamma - 1) l_2 + \tfrac{1}{2} \alpha l_3].
\tag{A21}
$$

and we have replaced the frequencies $\nu_i$ by their dimensionless counterparts $\tilde{\nu}_i$. After integration over energy, the integral in equation (A17) becomes

$$
\begin{aligned}
\omega_{\mathbf{l}} \int dI_1 J dJ |W_{l l_2 l_3}^{l_1}|^2 F_1(E, J) \delta(\mathbf{l} \cdot \boldsymbol{\Omega} - \omega_{\mathbf{l}}) = \\
-\frac{2 F_0 J_s^{2\gamma}}{e^\gamma} \frac{G^2 M_d^2}{R_d^2} \operatorname{sgn}(\omega_{\mathbf{l}}) \int_0^1 d\kappa \frac{\kappa^{2(\gamma - 1)}}{\tilde{\nu}_1(\kappa)} f(\kappa) |\widetilde{W}_{l l_2 l_3}^{l_1}(\kappa)|^2 \Theta\left(\frac{\mathbf{l} \cdot \tilde{\boldsymbol{\nu}}}{\tilde{\omega}_{\mathbf{l}}}\right),
\end{aligned}
\tag{A22}
$$

where the dimensionless transform is (*cf.* eq. A4)

$$
\widetilde{W}_{l l_2 l_3}^{l_1}(\kappa, \tilde{\omega}_{\mathbf{l}}) = \frac{1}{\pi} \int_{\tilde{r}_p}^{\tilde{r}_a} d\tilde{r} \frac{dw_1}{d\tilde{r}} \cos[l_1 w_1 - l_2(\psi - w_2)] \widetilde{U}_{l l_3}\left(\frac{\tilde{r} \mathbf{l} \cdot \tilde{\boldsymbol{\nu}}}{\tilde{\omega}_{\mathbf{l}}}\right),
\tag{A23}
$$

$\tilde{\omega}_{\mathbf{l}} = R_d \omega_{\mathbf{l}} / v_c$, and $U_{l l_3}(r) = (G M_d / R_d) \widetilde{U}_{l l_3}(r / R_d)$.

Finally, inserting this result in equation (A17) and using equation (71) we obtain the form we use for numerical computations,

$$\dot{E}_w = \frac{GM_d^2 v_c}{R_d^2} \dot{\mathcal{E}}(\widetilde{\Omega}_p; \gamma, \alpha), \tag{A24}$$

where

$$\dot{\mathcal{E}}(\widetilde{\Omega}_p; \gamma, \alpha) = \sum_{l_3=0}^{\infty} \sum_{l_1, l_2=-\infty}^{\infty} \sum_{l=1}^{\infty} \frac{4\pi}{(2l+1)} |Y_{ll_2}(\tfrac{1}{2}\pi, 0)|^2 \dot{\mathcal{E}}_{ll_2l_3}^{l_1}(\widetilde{\Omega}_p; \gamma, \alpha), \tag{A25}$$

and

$$\dot{\mathcal{E}}_{ll_2l_3}^{l_1}(\widetilde{\Omega}_p; \gamma, \alpha) = -\frac{\operatorname{sgn}(\widetilde{\Omega}_p)}{(1+\delta_{l_30})^2} \frac{4\pi^{1/2}\gamma^{\gamma+1/2}}{e^{\gamma}\,\Gamma(\gamma)} \int_0^1 d\kappa \frac{\kappa^{2(\gamma-1)}}{\widetilde{\nu}_1(\kappa)} f(\kappa) |\widetilde{W}_{ll_2l_3}^{l_1}(\kappa, l_3\widetilde{\Omega}_p)|^2 \Theta\left(\frac{\mathbf{l}\cdot\widetilde{\boldsymbol{\nu}}}{l_3\widetilde{\Omega}_p}\right). \tag{A26}$$

When $\gamma = 1$ and $\alpha = 0$, the DF is that of a singular isothermal sphere. In this case, we recover the result of Weinberg (1986; eq. 40), after correcting two minor typographical errors (the factor $2l+2$ in Weinberg's paper should be $2l+1$; and the sum over $l_3$ should run from 0 to $l$, not $-l$ to $l$) and converting from torque to energy loss.

Note the symmetry

$$\dot{\mathcal{E}}_{ll_2l_3}^{l_1}(\widetilde{\Omega}_p; \gamma, \alpha) = \dot{\mathcal{E}}_{l,-l_2l_3}^{-l_1}(-\widetilde{\Omega}_p; \gamma, -\alpha), \tag{A27}$$

which implies that

$$\dot{\mathcal{E}}(\widetilde{\Omega}_p; \gamma, \alpha) = \dot{\mathcal{E}}(-\widetilde{\Omega}_p; \gamma, -\alpha). \tag{A28}$$

Thus if the halo is non-rotating ($\alpha = 0$) and $l_3 \neq 0$, $\dot{E}_w$ is independent of the sign of the pattern speed as one would expect from symmetry.

The term in equation (A21) proportional to the halo spin $\alpha$ contributes energy to the bending wave if the pattern speed is in the same direction as the halo rotation, and removes energy if the pattern speed and halo rotation are opposite.

## B  Gravitational Potential of a Disk Bending Mode

The potential of a thin, warped disk of surface density $\mu(R)$ is

$$U_d(\mathbf{r}, t) = -G \int d\mathbf{R}' \frac{\mu(R')}{|\mathbf{r} - \mathbf{r}'(\mathbf{R}', t)|} \tag{B1}$$

where $\mathbf{r}' = \mathbf{R}' + Z(\mathbf{R}', t)\mathbf{e}_z$. Subtracting off the potential of the unperturbed (flat) disk, and expanding to first order in the small displacement $Z$, we obtain the potential due to the warp

$$U_w(\mathbf{r}, t) = -G \int d\mathbf{R}' \mu(R') Z(\mathbf{R}', t) \frac{\partial}{\partial z'} \frac{1}{|\mathbf{r} - \mathbf{r}'|}\bigg|_{z'=0}. \tag{B2}$$

In spherical coordinates

$$\frac{1}{|\mathbf{r} - \mathbf{r'}|} = \sum_{ln} \frac{4\pi}{2l+1} \frac{r_<^l}{r_>^{l+1}} Y_{ln}(\theta, \phi) Y_{ln}^*(\theta', \phi'), \tag{B3}$$

so that

$$\left. \frac{\partial}{\partial z'} \frac{1}{|\mathbf{r} - \mathbf{r'}|} \right|_{z'=0} = \sum_{ln} \frac{4\pi}{2l+1} \frac{r_<^l}{r_>^{l+1}} Y_{ln}(\theta, \phi) \frac{1}{R'} \frac{\partial Y_{ln}^*}{\partial \cos\theta} (\tfrac{1}{2}\pi, 0) e^{-in\phi'}. \tag{B4}$$

The values of the spherical harmonics are

$$\frac{\partial Y_{ln}}{\partial \cos\theta} (\tfrac{1}{2}\pi, 0) = (-1)^{(l+n-1)/2} \left( \frac{2l+1}{4\pi} \right)^{1/2} \sqrt{\frac{(l-n)!}{(l+n)!}} \frac{(l+n)!!}{(l-n-1)!!} \qquad n+l \quad \text{odd} \tag{B5}$$

where the double factorial terms are

$$(2n-1)!! = \frac{(2n)!}{2^n n!}, \qquad (2n)!! = 2^n n!. \tag{B6}$$

Terms with $l + n$ even vanish.

We now substitute equation (7) for $Z(\mathbf{R}, t)$. Then we may write

$$U(\mathbf{r}, t) = \sum_{l,n} U_{ln}(r) Y_{ln}(\theta, 0) e^{i(n\phi - \omega_n t)}, \tag{B7}$$

where

$$U_{ln}(r) = -\frac{4\pi^2 G}{2l+1} \frac{\partial Y_{lm}^*}{\partial \cos\theta} (\tfrac{1}{2}\pi, 0) \int_0^\infty dR' \mu(R') \frac{r_<^l}{r_>^{l+1}} [h(R')\delta_{mn} + h^*(R')\delta_{m,-n}], \tag{B8}$$

and $\omega_m = \omega$, $\omega_{-m} = -\omega^*$.

In this paper we mostly consider disturbances with $m = 1$. Since $n = \pm m$ and $n + l$ must be odd, only terms with even $l$ will be non-zero. We assume that the disk surface density and warp shape are given by equations (77) and (79). Then equation (B8) can be evaluated as

$$U_{l\pm 1}(r) = -\frac{GM_d}{R_d} \frac{2\pi}{2l+1} \frac{\partial Y_{l\pm 1}^*}{\partial \cos\theta} (\tfrac{1}{2}\pi, 0) \int_0^\infty dx f(x) \frac{x_<^l}{x_>^{l+1}} e^{-x} \tag{B9}$$

where

$$\int_0^\infty dx f(x) \frac{x_<^l}{x_>^{l+1}} e^{-x} = C \left[ \left( \frac{R_d}{r} \right)^{l+1} \frac{\gamma(n+l+1, (1+\lambda)r/R_d)}{(1+\lambda)^{n+l+1}} \right.$$
$$\left. + \left( \frac{r}{R_d} \right)^l \frac{\Gamma(n-l, (1+\lambda)r/R_d)}{(1+\lambda)^{n-l}} \right]. \tag{B10}$$

In this equation the gamma functions are defined by

$$\gamma(p, z) = \int_0^z x^{p-1} e^{-x} dx, \qquad \Gamma(p, z) = \int_z^\infty x^{p-1} e^{-x} dx; \tag{B11}$$

note that

$$\gamma(p,z) + \Gamma(p,z) = \Gamma(p) = (p-1)! \tag{B12}$$

If $p$ is a positive integer then

$$\Gamma(p,z) = (p-1)! e^{-z} \sum_{k=0}^{p-1} \frac{z^k}{k!}. \tag{B13}$$

For negative integers, $-n \leq 0$,

$$\Gamma(-n,z) = \frac{1}{z^n} E_{n+1}(z), \tag{B14}$$

where $E_n(z)$ is an exponential integral.

## C   The Contribution of Near-Radial Orbits to Dynamical Friction

In this Appendix we elucidate the reason for the apparent divergence in the damping/excitation rate in halos with predominantly radial orbits.

Suppose that a torque $\epsilon\boldsymbol{\tau}(t)$ is exerted on the star in a nearly radial orbit. The resulting change in the angular momentum vector may be written

$$\Delta \mathbf{J} = \epsilon \boldsymbol{\Delta}, \tag{C1}$$

where $\boldsymbol{\Delta} = \int \boldsymbol{\tau}(t) dt$. The change in $J = |\mathbf{J}|$ is then

$$\Delta J = \epsilon \frac{\mathbf{J} \cdot \boldsymbol{\Delta}}{J} + \epsilon^2 \left[ \frac{\boldsymbol{\Delta}^2}{2J} - \frac{(\mathbf{J} \cdot \boldsymbol{\Delta})^2}{2J^3} \right] + \mathrm{O}(\epsilon^3). \tag{C2}$$

As $J \to 0$, in general $\boldsymbol{\Delta}$ approaches a fixed, non-zero value. Thus as $J \to 0$

$$\Delta J \to \epsilon \Delta \cos \psi + \frac{\epsilon^2 \Delta^2 \sin^2 \psi}{2J}, \tag{C3}$$

where $\psi$ is the angle between $\boldsymbol{\Delta}$ and $\mathbf{J}$ (plus terms that are $O(\epsilon^3)$, which we neglect). The $\mathrm{O}(\epsilon)$ term can be dropped, since it averages to zero in a spherically symmetric distribution of stars, and in any case does not diverge as $J \to 0$. In a spherical distribution, the average of $\sin^2 \psi$ is $\frac{1}{2}$, so the average change is $\langle \Delta J \rangle = \frac{1}{4} \epsilon^2 \Delta^2 / J$. The corresponding energy change is

$$\langle \Delta E \rangle = \frac{\partial H_0}{\partial J} \langle \Delta J \rangle = \Omega_2 \langle \Delta J \rangle, \tag{C4}$$

which also diverges as $J^{-1}$, since $\Omega_2$ is generally constant and non-zero as $J \to 0$.

The number of stars in a small interval of energy and angular momentum is $dN = 16\pi^3 F(E,J) dE\, J dJ / \Omega_1$. Now assume that the DF has the form (70), that is $F(E,J) = J^{2(\gamma-1)} g(E)$. Since $\Omega_1$ is constant as $J \to 0$, the total energy change of the

– 36 –

stars of a given energy on near-radial orbits is proportional to $\int J^{2\gamma-1}\langle\Delta J\rangle dJ = \frac{1}{4}\epsilon^2\Delta^2\int J^{2\gamma-2}dJ$, which diverges if $\gamma \leq \frac{1}{2}$.

Thus we have explained the apparent divergence in dynamical friction for DFs with $\gamma \leq \frac{1}{2}$. Our arguments also show that the divergence is an artifact of the expansion procedure: the exact change in angular momentum is

$$\Delta J_{\text{exact}} = (J^2 + 2\epsilon\mathbf{J}\cdot\boldsymbol{\Delta} + \epsilon^2\Delta^2)^{1/2} - J, \tag{C5}$$

and the total energy or angular momentum change from near-radial orbits ($J < J_{\text{max}}$, say) is proportional to

$$\int_0^{J_{\text{max}}} J^{2\gamma-1}\langle\Delta J_{\text{exact}}\rangle dJ. \tag{C6}$$

When $\gamma > \frac{1}{2}$ this integral is $\text{O}(\epsilon^2)$ (after averaging over orientations assuming $\langle\cos\psi\rangle = 0$); this is the usual case in which the frictional energy or angular momentum change is second-order in the perturbation strength $\epsilon$. When $\gamma < \frac{1}{2}$ the integral is $\text{O}(\epsilon^{2\gamma+1})$; thus the frictional force is of a lower order in the perturbation strength than $\epsilon^2$, but is not divergent.

## Authors' Addresses

Robert W. Nelson and Scott Tremaine
CITA, University of Toronto
60 St. George Street
Toronto, ON, M5S 1A7
CANADA
Email: nelson@cita.utoronto.ca, tremaine@cita.utoronto.ca